\documentclass[11pt]{article}
 
\usepackage[utf8]{inputenc}
\usepackage{amsmath,amssymb}
\usepackage{bbold}
\usepackage{multirow,booktabs}
\usepackage[table]{xcolor}
\usepackage{fullpage}
\usepackage{slashed}
\usepackage{lastpage}
\usepackage{enumitem}
\usepackage{fancyhdr}
\usepackage[normalem]{ulem}
\usepackage{wrapfig}
\usepackage{setspace}
\usepackage{calc}
\usepackage{multicol}
\usepackage{cancel}
\usepackage[retainorgcmds]{IEEEtrantools}
\usepackage[margin=2cm]{geometry}
\usepackage[linkcolor={blue},citecolor={red},colorlinks=true]{hyperref}

\hypersetup{
    colorlinks=true,
    linkcolor=blue!60!black,
    citecolor=red
}

\usepackage{fontawesome5}

\usepackage{cleveref}
\usepackage{cancel}
\usepackage[]{cite}

\usepackage{empheq}
\usepackage{framed}
\usepackage[most]{tcolorbox}
\usepackage{xcolor}
\colorlet{shadecolor}{orange!15}

\def\bea{\begin{eqnarray}}
\def\eea{\end{eqnarray}}

\numberwithin{equation}{section}

\definecolor{chromeyellow}{rgb}{1.0, 0.65, 0.0}
\definecolor{darkcoral}{rgb}{0.8, 0.36, 0.27}
\definecolor{cadmiumgreen}{rgb}{0.0, 0.42, 0.24}

\begin{document}

\thispagestyle{empty}

\vspace{.6cm}
\begin{center}

{\Large \bf 
{Globally Charged Vacuum Decay}}
\vspace{0.5cm}

\vspace{1cm}{
Giulio Barni$^1$ and Jos\'e R. Espinosa$^2$
}
\\[7mm]
 {\it \small
Instituto de F\'isica Te\'orica IFT-UAM/CSIC, Cantoblanco, E-28049, Madrid, Spain \\[0.1cm]
}
\end{center}

\bigskip \bigskip \bigskip

\centerline{\bf Abstract} 
  \begin{quote}
Vacuum decay at zero temperature is generically described by a real $O(4)$-symmetric Coleman bounce. When the scalar field driving the decay 
carries a conserved global charge, this picture changes qualitatively: the path integral must be projected onto a definite charge sector, the Euclidean field obeys twisted boundary conditions, and the saddle is complex. For the simplest case of a $U(1)$ global symmetry, we first reformulate this problem in a two-field real Euclidean description with a real saddle.
We then solve the resulting two-dimensional partial differential equation problem describing the decay of a homogeneous charged medium to a deeper vacuum via bubble nucleation. At finite charge the bounce departs from $O(4)$ symmetry, the barrier between vacua is lowered, and the decay rate increases. Continuing the solution to real time, we find that charge rearrangement around the expanding wall costs phase-gradient energy and drives the bubble to a subluminal terminal velocity even in vacuum. We also clarify how the fixed-charge construction interfaces with finite-temperature and finite-chemical-potential descriptions.

\end{quote}
\vfill
\noindent\line(1,0){188}
{\scriptsize{ \\ E-mail:
\texttt{$^1$\href{giulio.barni@NOSPAMift.csic.es}{giulio.barni@ift.csic.es}},
\texttt{$^2$\href{jr.espinosa@NOSPAMcsic.es}{jr.espinosa@csic.es}}
}}
\newpage
\tableofcontents

\section{Introduction and motivations}

Semiclassical vacuum decay in relativistic field theory is usually formulated for neutral metastable states, where the dominant tunnelling channel is described by a real $O(4)$-symmetric Coleman bounce in Euclidean time \cite{Coleman:1977py,PhysRevD.16.1762}. This framework underlies the standard treatment of false-vacuum decay at zero temperature and its well-known extensions to gravity and finite temperature \cite{Coleman:1977py,PhysRevD.16.1762,Coleman:1980aw,Linde:1980tt,Linde:1981zj,Affleck:1981bma}. In that standard picture, the tunnelling problem is organised entirely in terms of an unconstrained order-parameter bounce.

The main subject of this paper is the semiclassical nucleation of bubbles out of a homogeneous metastable state at fixed conserved global charge. Our goal is to formulate this problem in a systematic way, solve the Euclidean partial differential equation (PDE) problem explicitly, compute the fixed-$Q$ decay exponent, and clarify the posterior Minkowski evolution. 

However, at finite density, the textbook framework does not apply if a conserved charge is present. In that case, the tunnelling event must occur inside a definite charge sector, and the semiclassical saddle is no longer the ordinary neutral bounce \cite{Lee:1994bza}. This issue is physically relevant whenever a medium carries a conserved density and charge exchange with the environment is inefficient on the timescale of nucleation. In such situations a chemical potential may provide an effective macroscopic description, but the microscopic tunnelling event itself should still be treated at fixed charge. This viewpoint is especially natural for phase conversion problems in dense relativistic matter and compact-stars  \cite{Glendenning:1992vb,Iida:1997pj,Mintz:2009ay,Dexheimer:2018dlz,Bauswein:2018bma,Cao:2018tzm,Casalderrey-Solana:2022rrn}.

We formulate the fixed-charge tunnelling problem in a real Euclidean language adapted to constrained semiclassics, together with a conjugate twist parameter enforcing charge projection. In the simplest $U(1)$ case studied here, this takes the explicit form of a doubled-field formulation in terms of two independent Euclidean fields. This makes the variational problem real and numerically tractable, while preserving the intrinsically complex character of the underlying saddle. The resulting construction produces charged saddles that depart from $O(4)$ symmetry and lead to nontrivial real-time bubble dynamics.

More broadly, if one ultimately wants quantitative nucleation estimates in systems at finite charged density, the appropriate microscopic building block is not the ordinary neutral bounce and not a grand-canonical saddle imposed from the outset, but the tunnelling solution in a definite charge sector. The present analysis should therefore be viewed as a first fully worked-out step toward a quantitative semiclassical framework for phase conversion in systems with finite conserved charge density.

\subsection{Setup}

We thus focus on the problem of semiclassical tunnelling in relativistic field theory at zero temperature ($T=0$) when the tunnelling scalar field carries a conserved global $U(1)$ charge. The microscopic theory describes a complex scalar field with global $U(1)$ invariance and action
\begin{equation}
S=\int d^4x\,\left[\partial_\mu\Phi\,\partial^\mu\Phi^\ast - V(|\Phi|^2)\right],
\end{equation}
with associated Noether current and total charge
\begin{equation}
\label{eq:current}
j^\mu=i\Big(\Phi\partial^\mu\Phi^\ast-\Phi^\ast\,\partial^\mu\Phi\Big),
\qquad
Q=\int d^3x\,j^0.
\end{equation}

The physical picture we have in mind is as follows. Our initial state is a homogeneous charged state (\emph{i.e.} a spatially uniform configuration with finite charge density) which is classically stable.\footnote{One can think of these initial states as $Q$-ball states in the limit of infinite $Q$-ball radius. For the connection of our work to previous $Q$-ball analyses, see section \ref{sec:lit}.} However,  that state is quantum mechanically metastable due to the existence of another charged homogeneous state with lower energy, to which the initial state can decay through bubble nucleation.

A charged homogeneous equilibrium configuration is given by a stationary rotating state,
\begin{equation}
\Phi(t,\mathbf x)=\frac{\rho}{\sqrt2}\,e^{i\omega t},
\label{eq: initial_state}
\end{equation}
with constant $\rho$ and $\omega$. Such a state carries nonzero uniform charge density $q\equiv j^0=\omega\rho^2$, total charge $Q=V_3 \omega \rho^2$, and total energy $E=V_3[V(\rho^2/2)+\omega^2\rho^2/2]$, where $V_3$ is the three-dimensional volume of the box considered. Classical stability of such state requires 
\begin{equation}
    V'(\rho^2/2)-\omega^2=0\ ,
    \label{classtab}
\end{equation}
which follows from $\Phi$'s equation of motion (EoM). We therefore assume that our potential is such that (\ref{classtab}) has two solutions: the initial metastable state $\Phi_i=(\rho_i/\sqrt{2})e^{i\omega_i t}$ and the final stable state $\Phi_f=(\rho_f/\sqrt{2})e^{i\omega_f t}$. After full phase conversion, charge conservation would require
$\rho_i^2 \omega_i=\rho_f^2\omega_f$. We will also assume $\rho_f>\rho_i$, which implies $\omega_f<\omega_i$. The condition that $\Phi_f$ has lower energy than $\Phi_i$ implies $V(\rho_f^2/2)<V(\rho_i^2/2)-q(\omega_i-\omega_f)/2$.

Decay of the metastable state occurs by bubble nucleation as in the standard $Q=0$ case, but the Euclidean bounce techniques cannot be applied directly. 
The root of the difficulty is not that the charged field is necessarily complex (it could always be decomposed in two real fields) but rather that the
metastable state is time dependent and going to Euclidean time ($t\to -i\tau$) makes the Euclidean saddle complex, as we will see.

Our objective is twofold. First, we formulate the fixed-$Q$ decay exponent in a controlled semiclassical expansion, making explicit the differences with both standard Coleman tunnelling at $Q=0$ \cite{Coleman:1977py,PhysRevD.16.1762} and metastable $Q-$ball decay \cite{Levkov:2017paj}. Second, we solve the resulting Euclidean PDE problem numerically, identify the physically relevant branch among several nontrivial solutions, and reconstruct the post-tunnelling Minkowski evolution. In this sense the paper is not only conceptual but computational: we solve the full fixed-$Q$ saddle problem and compute the decay exponent explicitly.\footnote{All numerical material and calculations presented in this work are publicly available on the GitHub page \href{https://github.com/GiulioBarni/Qubble}{\faGithub\ Qubble}, which also contains the full reproduction of the results of Ref.~\cite{Levkov:2017paj}.}

In doing so, we will find three main effects. First, the decay exponent is reduced relative to the neutral case, so the decay rate is enhanced at finite charge. Physically, this reflects the fact that the conserved charge lowers the effective potential barrier, making bubble nucleation easier than in the uncharged vacuum. Second, the charge deforms the Euclidean saddle away from the standard $O(4)$ Coleman geometry, so that the charged bounce is a function of the two variables $r$ and $\tau$ separately. Third, the real-time evolution after nucleation is qualitatively modified: the bubble wall does not accelerate all the way to the speed of light, but instead approaches a terminal velocity smaller than one. Physically, this happens because the bubble expands inside a charged medium, and the conserved charge must continuously readjust during the evolution. The associated phase gradients build up around the moving wall, contribute to its energy, and eventually stop the runaway wall acceleration.

This work is organised as follows. In Sec.~\ref{sec:quantum_tunnelling_Q} we
formulate the tunnelling problem in a fixed-charge sector, introduce the charge
projection in the path integral, and derive the corresponding twisted
Euclidean-time boundary conditions. We then present the real Euclidean
formulation in terms of independent fields, discuss the associated conservation
laws, and compare the setup with previous approaches. In
Sec.~\ref{sec:SQbeta} we derive the semiclassical fixed-charge tunnelling
exponent, clarify its relation to a modified Euclidean action and give an
energy-barrier interpretation. In Sec.~\ref{sec:example} we illustrate the
formalism in an explicit model and present the numerical bounce solutions
together with the main consistency checks, including the
$\beta\equiv 1/T\to\infty$ and $Q\to0$ limits. In
Sec.~\ref{sec:thin_wall_fixed_charge} we develop the thin-wall approximation at
fixed charge, which includes a non-trivial phase profile. We calculate the
finite-$\beta$ action, the decay-rate estimate, and the modulus response
required by energy conservation. In Sec.~\ref{sec:Minkowski} we discuss the
analytic continuation to real time and the Lorentzian evolution of the
nucleated bubble, with particular emphasis on late-time wall energetics and
terminal-velocity estimates. In Sec.~\ref{sec:extension} we comment on the
extension of the formalism to finite temperature and finite chemical potential.
Finally, in Sec.~\ref{sec:conclusion} we summarise our results and outline
possible directions for future work. Appendix~A contains some details on how
the charged field phase affects the late-time wall dynamics.

\section{Quantum tunnelling at $Q\neq0$}
\label{sec:quantum_tunnelling_Q}

As we have seen, for our discussion the relevant metastable state is a charged homogeneous configuration rather than a neutral false vacuum. The tunnelling event must take place within a definite charge sector, so the semiclassical problem is not the ordinary Coleman problem in a different background, but a constrained variational problem.
Two structural differences with respect to the neutral case are the following. First, the fixed-$Q$ path integral is obtained by a projection in a charged sector, and this introduces a variable conjugate to $Q$ and manifests itself as a twist in Euclidean-time boundary conditions. Second, the saddle solution is not represented by a single Euclidean field and its complex conjugate; rather, the semiclassical formulation treats two Euclidean fields, $\phi$ and $\bar\phi$, as independent variables. We review this below, following \cite{Levkov:2017paj} closely. For earlier literature on this, see section~\ref{sec:lit}.

\subsection{Fixed-$Q$ partition function and charge projection}

The natural canonical object is the partition function restricted to a definite charge sector,
\begin{equation}
Z_Q(\beta)\equiv \mathrm{Tr}_Q\!\left(e^{-\beta \hat H}\right),
\label{eq:ZQdef}
\end{equation}
where $\mathrm{Tr}_Q$ denotes the trace over states of total charge $Q$ and, for later convenience, we take the Euclidean time interval to be $(-\beta/2,\beta/2)$. In the large-$\beta$ limit, this projects onto the lowest-energy state in that sector \cite{Lee:1988ge, Lee:1994bza},
\begin{equation}
Z_Q(\beta)\sim e^{-\beta E_0(Q)},
\qquad
\beta\to\infty.
\label{eq:ZQlargebeta}
\end{equation}
To enforce the restriction to fixed charge, one inserts a projector onto the charge-$Q$ sector. Formally one may write
\begin{equation}
\hat P_Q=\delta(\hat Q-Q),
\label{eq:PQdelta}
\end{equation}
and represent the delta function in Fourier form as
\begin{equation}
\hat P_Q
=
\int_0^{2\pi}\frac{d\theta}{2\pi}\,
e^{i\theta(\hat Q-Q)}.
\label{eq:PQ}
\end{equation}
The compact interval reflects the fact that $e^{2\pi i\hat Q}=1$ for integer charge eigenvalues.\footnote{
Strictly speaking, the charge is quantised: $Q = n Q_0$ with $n \in \mathbb{Z}$, due to the $U(1)$ symmetry, and the projector $\hat P_Q$ selects a fixed integer sector. In general, one works at fixed $n$ or sums over sectors. However, for the semiclassical cases of interest, $n \gg 1$,  the discrete spectrum can be treated as continuous and, in this limit, it is safe to treat $Q$ as a continuous variable.
} 
Inserting \eqref{eq:PQ} into the trace gives
\begin{equation}
Z_Q(\beta)
=
\mathrm{Tr}\!\left(\hat P_Q e^{-\beta \hat H}\right)
=
\int_0^{2\pi}\frac{d\theta}{2\pi}\,
e^{-i\theta Q}\,
\mathrm{Tr}\!\left(e^{-\beta \hat H+i\theta\hat Q}\right).
\label{eq:ZQprojected}
\end{equation}
Equation \eqref{eq:ZQprojected} is the starting point of the fixed-$Q$ path-integral formulation.

\subsection{Path integral with twisted Euclidean-time boundary conditions}

To convert \eqref{eq:ZQprojected} into a path integral, one inserts complete sets of field eigenstates in the trace. Let $\hat\Phi(\mathbf x)$ be the complex scalar field operator. The operator $e^{i\theta\hat Q}$ implements a global $U(1)$ rotation,
\begin{equation}
e^{i\theta\hat Q}\,\hat\Phi(\mathbf x)\,e^{-i\theta\hat Q}
=
e^{i\theta}\hat\Phi(\mathbf x),
\qquad
e^{i\theta\hat Q}\,\hat\Phi^\dagger(\mathbf x)\,e^{-i\theta\hat Q}
=
e^{-i\theta}\hat\Phi^\dagger(\mathbf x).
\label{eq:U1rotation}
\end{equation}
Accordingly, if $|\phi\rangle$ is an eigenstate of $\hat\Phi(\mathbf x)$, that is 
$
\hat\Phi(\mathbf x)|\phi\rangle=\phi(\mathbf x)|\phi\rangle,
$
then $e^{i\theta\hat Q}|\phi\rangle$ is an eigenstate with eigenvalue $e^{i\theta}\phi(\mathbf x)$. Indeed,
\begin{equation}
\hat\Phi(\mathbf x)\,e^{i\theta\hat Q}|\phi\rangle
=
e^{i\theta\hat Q}\Bigl(e^{-i\theta\hat Q}\hat\Phi(\mathbf x)e^{i\theta\hat Q}\Bigr)|\phi\rangle
=
e^{i\theta}\phi(\mathbf x)\,e^{i\theta\hat Q}|\phi\rangle,
\end{equation}
so, up to an irrelevant phase convention, $e^{i\theta\hat Q}|\phi\rangle = |e^{i\theta}\phi\rangle$.

Inserting a complete set of field eigenstates into the trace then gives
\begin{equation}
{\rm Tr}\!\left(e^{-\beta \hat H}e^{i\theta \hat Q}\right)
=
\int \mathcal D\Phi\,\mathcal D\Phi^\ast\;
\langle \phi|e^{-\beta \hat H}e^{i\theta \hat Q}|\phi\rangle
=
\int \mathcal D\Phi\,\mathcal D\Phi^\ast\;
\langle \phi|e^{-\beta \hat H}|e^{i\theta}\phi\rangle.
\end{equation}
Therefore, in the present case, the initial and final configurations are identified only up to a global $U(1)$ rotation, with the twisted boundary 
conditions
\begin{equation}
\Phi(\beta/2,\mathbf x)=e^{i\theta}\Phi(-\beta/2,\mathbf x),
\qquad
\Phi^\ast(\beta/2,\mathbf x)=e^{-i\theta}\Phi^\ast(-\beta/2,\mathbf x).
\label{eq:twistedBCtheta}
\end{equation}
As a result, \eqref{eq:ZQprojected} becomes
\begin{equation}
Z_Q(\beta)
=
\int_0^{2\pi}\frac{d\theta}{2\pi}\,e^{-i\theta Q}
\int
\mathcal D\Phi\,\mathcal D\Phi^\ast\;
e^{-S_E[\Phi,\Phi^\ast;\theta]}\ ,
\label{eq:ZQtwistedPI}
\end{equation}
where the twisted boundary conditions (\ref{eq:twistedBCtheta}) are assumed.
This is the basic mechanism by which fixed charge enters the Euclidean saddle problem \cite{Levkov:2017paj}.
The key point is that the dominant saddle of the $\theta$ integral does not in general lie at real $\theta$. Instead, steepest descent selects an imaginary twist,
\begin{equation}
\theta=-i\eta,
\qquad
\eta\in\mathbb R\ .
\label{eq:thetaeta}
\end{equation}
This is directly tied to the reality of the charge. If, after Wick rotation, one insisted on keeping an ordinary real phase, then the Euclidean continuation of \eqref{eq:current} would give a purely imaginary charge. In other words, a real Euclidean phase would describe an imaginary charge density. This is not the correct saddle for a tunnelling process at fixed real charge. For more details, see \cite{Barni:2026dhc}.

Thus the boundary conditions become
\begin{equation}
\Phi(\beta/2,\mathbf x)=e^{\eta}\Phi(-\beta/2,\mathbf x),
\qquad
\Phi^\ast(\beta/2,\mathbf x)=e^{-\eta}\Phi^\ast(-\beta/2,\mathbf x).
\label{eq:twistedBCeta}
\end{equation}
These are not periodic boundary conditions with a phase, but quasi-periodic boundary conditions with a real exponential factor. The parameter $\eta$ is conjugate to the charge in the projected path integral, and its saddle value $\eta_b$ is fixed so that the semiclassical configuration carries the desired total charge, see below.

As a consequence, the saddle can no longer be described in terms of a single field and its ordinary complex conjugate. One can instead use the doubled-field formulation, treating
\begin{equation}
\phi(\tau,\mathbf x),
\qquad
\bar\phi(\tau,\mathbf x),
\end{equation}
as independent Euclidean fields obeying the quasi-periodic boundary conditions \eqref{eq:twistedBCeta}.

\subsection{Real Euclidean formulation}
\label{sec:euclidean}

The appropriate semiclassical problem is therefore formulated in the complexified field space, where the two Euclidean fields are treated as independent variables. A convenient real section of this complexified problem is obtained by introducing two real Euclidean fields $\phi$ and $\bar\phi$ and considering the action
\begin{equation}
S_E[\phi,\bar\phi]
=
\int_{-\beta/2}^{\beta/2} d\tau \int d^3x\,
\left[
\partial_\tau\phi\,\partial_\tau\bar\phi
+
\nabla\phi\cdot\nabla\bar\phi
+
V(\phi\bar\phi)
\right].
\label{eq:SEphibarphi}
\end{equation}
For real $\phi$ and $\bar\phi$, the functional \eqref{eq:SEphibarphi} is manifestly real, making the problem accessible to a real numerical treatment, even though the underlying saddle, viewed in the original Minkowski variables, is complex.

Varying \eqref{eq:SEphibarphi} with respect to $\phi$ and $\bar\phi$ gives the coupled Euclidean equations
\begin{equation}
(\partial_\tau^2+\nabla^2)\phi
-
V'(\phi\bar\phi)\,\phi
=
0,
\qquad
(\partial_\tau^2+\nabla^2)\bar\phi
-
V'(\phi\bar\phi)\,\bar\phi
=
0,
\label{eq:EOM_full}
\end{equation}
where $V'$ denotes the derivative of $V$ with respect to its argument.
Assuming spherical symmetry in space, the fields depend only on $(r,\tau)$ with $r=|\mathbf{x}|$, and \eqref{eq:EOM_full} gives
\begin{equation}
\partial_\tau^2\phi
+
\partial_r^2\phi
+
\frac{2}{r}\partial_r\phi
-
V'(\phi\bar\phi)\phi
=
0,
\label{eq:EOM_phi_O3}
\end{equation}
\begin{equation}
\partial_\tau^2\bar\phi
+
\partial_r^2\bar\phi
+
\frac{2}{r}\partial_r\bar\phi
-
V'(\phi\bar\phi)\bar\phi
=
0.
\label{eq:EOM_phibar_O3}
\end{equation}

For the boundary conditions, at the origin, regularity requires
\begin{equation}
\partial_r\phi(\tau,0)=0,
\qquad
\partial_r\bar\phi(\tau,0)=0.
\label{eq:regularity_r0}
\end{equation}
At large radius, the bounce must approach the initial homogeneous charged background,
\begin{equation}
\phi(\tau,r)\xrightarrow[r\to\infty]{}\phi_i(\tau),
\qquad
\bar\phi(\tau,r)\xrightarrow[r\to\infty]{}\bar\phi_i(\tau).
\label{eq:large_r_BC}
\end{equation}
On the full Euclidean interval, the fixed-charge projection imposes the twisted closure
\begin{equation}
\phi(\beta/2,r)=e^{\eta_b}\phi(-\beta/2,r),
\qquad
\bar\phi(\beta/2,r)=e^{-\eta_b}\bar\phi(-\beta/2,r).
\label{eq:twisted_closure_full}
\end{equation}
With our convention for the Euclidean time interval the turning slice occurs at $\tau=0$. At this surface Euclidean data are matched to physical Minkowski initial conditions. The turning-point conditions are
\begin{equation}
\phi(0,r)=\bar\phi(0,r),
\qquad
\partial_\tau\phi(0,r)=-\partial_\tau\bar\phi(0,r).
\label{eq:turning_conditions}
\end{equation}
The symmetries of the saddle equations imply the reflection-exchange symmetry \cite{Levkov:2017paj} 
\begin{equation}
\phi(\tau,r)=\bar\phi(-\tau,r),
\qquad
\bar\phi(\tau,r)=\phi(-\tau,r),
\label{eq:reflection_exchange}
\end{equation}
so that the branch at $\tau>0$ is reconstructed from the solution on $\tau<0$.\footnote{Thus the Euclidean action is twice the half-interval action,
$S_E[\phi,\bar\phi]=2S_E^{\rm half}[\phi,\bar\phi]$,
provided the turning conditions hold.}
As in the standard bounces, $\tau=0$
is the ``turning slice" which gives the nucleated bubble and the $\tau>0$ evolution of the bounce is a mirror copy of the $\tau<0$ evolution.
For later use, one can combine (\ref{eq:twisted_closure_full}) and (\ref{eq:reflection_exchange}) to get
\begin{align}
    \phi(-\beta/2,r)=e^{-\eta_b}\bar\phi(-\beta/2,r)\ ,\quad
    \phi(\beta/2,r)=e^{\eta_b}\bar\phi(\beta/2,r)\ .
    \label{eq:twisted_reflected}
\end{align}

Thus, in Euclidean time, $\phi$ and $\bar\phi$ are allowed to depart from conjugacy so as to implement the fixed-charge twist and to lie on the correct steepest-descent contour. Physical complex conjugation is recovered in real time, as the solution is analytically continued at the turning slice.

For later use it will be convenient to use a polar representation of the Euclidean fields as
\begin{equation}
\phi(r,\tau)=\frac{\rho(r,\tau)}{\sqrt{2}}\,e^{\alpha(r,\tau)}\ ,
\qquad
\bar\phi(r,\tau)=\frac{\rho(r,\tau)}{\sqrt{2}}\,e^{-\alpha(r,\tau)}\ .
\label{eq:polar_phi_phibar}
\end{equation}
The scalar potential $V(\phi\bar\phi)$ is a function of the modulus only, $V(\rho^2/2)$. With an abuse of notation we simply write $V(\rho)$ below, keeping in mind that $V'\equiv dV/d(\phi\bar\phi)$ in the bounce EoMs for $\phi$ and $\bar\phi$. 
In these polar coordinates, the Euclidean EoMs are 
\begin{align}
\partial_\mu(\rho^2\partial_\mu\alpha)=0\ ,\quad\quad
    \partial_\mu^2\rho  = \frac{\partial}{\partial\rho}\left[ V(\rho) -\frac12 \rho^2 (\partial_\mu\alpha)^2\right]\ ,
    \label{eq:polarEoMs}
\end{align}
where we have written the phase part as a contribution to a ``reduced potential", see discussions below.

\subsection{General energy and charge conservation}

In the quantum tunnelling process that connects the initial homogeneous charged configuration, $\Phi_i=(\rho_i/\sqrt{2})e^{i\omega_i t}$, and the nucleated bubble, $\Phi_B$, energy is conserved.
The energy of a field configuration $\Phi(\mathbf{x},t)$ is given by the functional
\begin{equation}
    E[\Phi]=\int d^3x \left[\partial_t\Phi^\dagger\partial_t\Phi + \nabla\Phi^\dagger\nabla\Phi +V(\Phi^\dagger\Phi)\right]\ .
\label{Energy}
\end{equation}
This quantity is conserved under standard Minkowski evolution for $\Phi$ satisfying its equation of motion. To prove energy conservation
during the Euclidean time evolution that describes tunnelling, we simply continue (\ref{Energy}) to Euclidean time, defining
\begin{equation}
    {\cal E}(\tau)=\int d^3x \left[-\partial_\tau\bar\phi\partial_\tau\phi + \nabla\bar\phi\nabla\phi +V(\bar\phi\phi)\right]\ .
\label{EucEnergy}
\end{equation}
It can be easily checked that this quantity is conserved, $\partial_\tau{\cal E}=0$, for $\bar\phi$ and $\phi$ satisfying their Euclidean EoMs (and regularity conditions). This conservation equation holds because these EoMs also follow from Euclidean continuation of the Minkowski ones.
The final part of the proof is to show that ${\cal E}(\tau\to\pm\infty)$ equals the energy of the initial homogeneous state while 
at the turning point, $\tau=0$, ${\cal E}(0)$ equals the energy of the nucleated bubble. For the $\tau\to\pm\infty$ field configuration we have 
the asymptotics (\ref{eq:phi_asymp}) and (\ref{eq:phi_asymp2}) so that
\begin{equation}
    {\cal E}(\pm\infty)=E[\Phi_i]=\int d^3x \left[\frac12\rho_i^2\omega_i^2 +V(\rho_i^2/2)\right]\ .
\end{equation}
For the nucleated bubble we have $\Phi_B=\phi(\mathbf{x},0)=\bar\phi(\mathbf{x},0)$ and $\partial_t\Phi_B=i\partial_\tau\phi(\mathbf{x},0)$, $\partial_t\Phi_B^\dagger=i\partial_\tau\bar\phi(\mathbf{x},0)$ so that ${\cal E}(0)=E[\Phi_B]$. So we conclude $E[\Phi_i]=E[\Phi_B]$, as needed.

The same logic can be used to show how charge conservation works in detail, with the definition of total charge extended to Euclidean time
as\footnote{Anticipating the conservation of this quantity throughout Euclidean and Minkowskian evolution, we use the same name $Q$ for
this Euclidean functional.}
\begin{equation}
    Q=\int d^3x\, (\bar\phi\partial_\tau \phi- \phi\partial_\tau \bar\phi)=\int d^3x\, \rho^2\, \partial_\tau\alpha\ .
\end{equation}
For $\phi$ and $\bar\phi$ satisfying the Euclidean EoMs, $Q$ is conserved during Euclidean time evolution, $dQ/d\tau=0$, and
$Q(\tau\to\pm\infty)=Q_i$ while $Q(0)=Q_B$, so that the total charge of the initial homogeneous state is equal to the total charge of the 
state containing the nucleated bubble, $Q_i=Q_B$. 

\subsection{Comparison to previous literature}
\label{sec:lit}

The basic Euclidean method for decay at fixed global charge is not new. Its roots go back to the wormhole literature \cite{Lee:1988ge,Coleman:1989zu,Giddings:1988cx,Giddings:1988wv,Grinstein:1988ja}, where charge projection, imaginary symmetry 
directions, and real formulations of complex saddles were developed in order to describe semiclassical processes at fixed conserved global charge. The first direct application of such approach to false-vacuum decay in the presence of a homogeneous charge density was then given by Lee \cite{Lee:1994bza}, who showed that the Euclidean phase becomes imaginary on the stationary path, that the bounce describing decay is no longer $O(4)$ symmetric, and that bubble growth after nucleation is correspondingly modified.\footnote{A broader literature has explored departures from the standard neutral Coleman picture, including induced vacuum decay, particle-catalysed decay, and tunnelling from  non-vacuum initial states (for an incomplete list of references, see \cite{Affleck:1981bma,Voloshin:1986zq,Voloshin:1990mz,Voloshin:1993ks,Gorsky:1993ix,Gorsky:2005yq,Dasgupta:1997kn,Lee:2013zca,
Blasi:2024mtc,Chatrchyan:2025uar,Steinhardt:1981ec,Yajnik:1986tg,Yajnik:1986wq,Blasi:2022woz,Agrawal:2023cgp,
Bai:2025qch,Blasi:2023rqi}). These works show that the $O(4)$-symmetric bounce is not the universal description once additional conserved quantities or nontrivial initial-state structure are involved. Finite-density scenarios provide further motivation: conserved charge densities or chemical potentials can qualitatively alter the fate of supercooled phases, for example, through charge supersaturation and chemical-potential-induced condensation \cite{Baratella:2025mwi}. }

A more recent work that is especially relevant to us is \cite{Levkov:2017paj}, which studies the decay of metastable $Q-$balls into their particle components. As in \cite{Lee:1994bza}, while the relevant saddle is complex in the original variables the decay exponent can be calculated through a real variational problem in terms of two independent Euclidean fields.\footnote{The bounce equations in \cite{Levkov:2017paj} are nothing but the Cartesian version of those in \cite{Lee:1994bza}, which are given in polar coordinates.} Our initial homogeneous charged state can also be thought of as a $Q$-ball, but we are considering it has infinite radius rather than the finite radius of the localised soliton sitting in vacuum of \cite{Levkov:2017paj}. Moreover, we are interested in the decay
by nucleation of a bubble of a phase (with lower energy density) which also supports $Q$ balls of infinite radii. Our boundary conditions are therefore different and our tunnelling saddle must approach a spatially uniform charged configuration at large radius.

The novelty of the present work therefore lies first in the different sort of problem we attack compared to \cite{Levkov:2017paj} and second,
in comparison to \cite{Lee:1994bza}, we make progress in presenting a systematic, powerful, and computation-ready form for bubble nucleation out of a homogeneous charged medium including in particular the possibility of addressing thick-wall cases.  In particular, no previous work had provided an explicit numerical construction of the charged Euclidean saddle, a finite and well-defined fixed-$Q$ suppression exponent relative to the homogeneous charged background, and a controlled reconstruction of the Minkowski evolution.

The problem also fits naturally within the recent effort to sharpen the relation between Euclidean saddles, complexified semiclassics, and real-time tunnelling dynamics \cite{Andreassen:2016cff,Andreassen:2016cvx,Steingasser:2023gde,Steingasser:2024ikl,Lin:2025bjn,Lin:2025wgc}. In this context, the fixed-$Q$ bounce provides a particularly clean laboratory: the tunnelling event is constrained, the Euclidean saddle is genuinely complex in the original variables, the corresponding real section can nevertheless be formulated explicitly, and the Minkowski continuation is physically nontrivial. More detailed and general results can be found in \cite{Barni:2026dhc}.  One of the main messages of this paper is precisely that finite charge modifies not only the exponential suppression, but also the post-nucleation wall dynamics, leading in particular to a terminal velocity smaller than one even at zero temperature.

\section{Semiclassical tunnelling exponent}
\label{sec:SQbeta}

The fixed-$Q$ path integral \eqref{eq:ZQtwistedPI} is now evaluated by steepest descent, both in field space and in the twist variable. The fixed-charge projector imposes the total quasiperiodicity in \eqref{eq:twistedBCeta}, and for a saddle satisfying this twisted closure, the projected path integral gives schematically
\begin{equation}
Z_Q(\beta)
\sim
\int d\eta\;
\exp\!\left\{
-\,S_E[\Phi_\eta,\Phi^\ast_\eta]-\eta Q
\right\},
\label{eq:ZQeta_schematic}
\end{equation}
where $(\Phi_\eta,\Phi^\ast_\eta)$ denote the Euclidean saddles satisfying the twisted closure at fixed $\eta$. The saddle-point condition in $\eta$ selects the configuration with the desired total charge, see below. At the dominant saddle $\eta=\eta_b$, the exponent is 
\begin{equation}
S^{\rm bounce}_{Q,\beta}
=
S_E[\Phi_{\eta_b},\Phi^\ast_{\eta_b}]
+\eta_b Q\ .
\label{eq:Fraw}
\end{equation}

For a metastable state, the decay rate is extracted from the nonperturbative
one-bounce contribution to the fixed-$Q$ partition function, normalised by the
zero-bounce contribution associated with the metastable homogeneous state. This
is the fixed-charge analogue of the dilute-instanton-gas argument in false
vacuum decay: the probability per unit time is controlled by the ratio\footnote{
Strictly speaking, the full fixed-$Q$ trace has no imaginary part:
when all states/sectors are included, probability is conserved and the would-be
imaginary contributions cancel. The decay rate is obtained by isolating
the contribution associated with the metastable basin, or equivalently by
computing the transition probability out of that basin. This is the standard
false-vacuum logic underlying the dilute-instanton-gas formula, and is the
viewpoint made explicit in Refs.~\cite{Levkov:2017paj,Steingasser:2023gde,
Steingasser:2024ikl,Lin:2025wgc,Lin:2025bjn}.
}
\begin{equation}
\Gamma \propto \frac{{\rm Im \ }Z_Q^{\rm bounce}}{Z_Q^{\rm fv}}
\sim
\exp[-S_{Q,\beta}]\ ,
\end{equation}
up to the usual prefactor and the overall spacetime volume factor. Thus the
relevant exponent is not the absolute saddle bounce action appearing in
\eqref{eq:ZQeta_schematic}, but the bounce contribution divided by the
metastable reference contribution (the false vacuum one) in the same charge sector. 
{Both configurations contain extensive contributions proportional to $\beta V_3$, which cancel in the nucleation exponent.

For the nontrivial bounce saddle one obtains
\begin{equation}
{\rm Im \ }Z_Q^{\rm bounce}
\sim
\exp\!\left\{
-S_E[\phi_b,\bar\phi_b]-\eta_bQ
\right\}\ ,
\end{equation}
whereas the false vacuum contribution is dominated, at large $\beta$, by the
metastable homogeneous state of energy $E_i(Q)$,
\begin{equation}
Z_Q^{\rm fv}
\sim
\exp[-\beta E_i(Q)].
\end{equation}
Therefore
\begin{equation}
S_{Q,\beta}
=
S_E[\phi_b,\bar\phi_b]
+\eta_bQ
-\beta E_i(Q)\ .
\label{eq:FQbeta_eta_cl}
\end{equation}

It is useful to rewrite this expression in terms of the Euclidean action of the
homogeneous charged reference configuration. For the Euclidean continuation of the homogeneous charged state one may take
\begin{equation}
\phi_i(\tau)={\rho_i \over \sqrt{2}} e^{+\omega_i\tau},
\qquad
\bar\phi_i(\tau)={\rho_i \over \sqrt{2}} e^{-\omega_i\tau},
\label{eq:hom_Euclidean_fields}
\end{equation}
so that $2\phi_i\bar\phi_i=\rho_i^2$ is constant. The homogeneous Minkowski energy is
\begin{equation}
E_i(Q)
=
\int d^3x
\left[|\dot\Phi_i|^2+V(|\Phi_i|^2)
\right]
=
V_3\left[V(\rho_i^2/2)+{1 \over 2}\omega_i^2\rho_i^2\right],
\label{eq:EhomQ}
\end{equation}
while the Euclidean action of the homogeneous configuration takes the form
\begin{equation}
S_E[\phi_i,\bar\phi_i]
=
\beta V_3\left[V(\rho_i^2/2)-{1 \over2 }\omega_i^2\rho_i^2\right]
=
\beta E_i(Q)-\eta_i Q\ ,
\label{eq:SEhom}
\end{equation}
where we have introduced the twist $\eta_i\equiv \beta\omega_i$. With ``twist'' we refer to the total change of the phase accumulated along the  Euclidean-time interval. Since the phase is an angular variable, fixing the charge is equivalent to fixing the variable conjugate to this total phase rotation. The twist for the homogeneous charged state is $\eta_i$ and for the bounce it is $\eta_b$. Their difference, 
\begin{equation}
\Delta\eta 
\equiv
\eta_b-\eta_i\ ,
\label{eq:eta0_eta_cl_relation}
\end{equation}
is the residual twist required to keep the bounce in the same fixed-charge sector as the initial state.\footnote{Geometrically, one can think of the complex field as an arrow in the internal complex plane. The twist is the total angle by which this arrow rotates as one moves from one end of the Euclidean-time interval to the other. The residual twist measures how much this total rotation must change when a bubble is inserted, while keeping the total charge fixed.}

Putting all this together, one obtains the fixed-$Q$ suppression exponent 
\begin{equation}
S_{Q,\beta}
=
S_E[\phi_b,\bar\phi_b]
-
S_E[\phi_i,\bar\phi_i]
+
\Delta\eta  Q\ , 
\label{eq:FQbeta_intro_new}
\end{equation}
where $(\phi_b,\bar\phi_b)$ is the nontrivial Euclidean saddle and $(\phi_i,\bar\phi_i)$ is the metastable reference homogeneous configuration in the same charge sector.
This is the form used below. It is also the form naturally implemented in the
numerical problem, where we work with fields rotated by the homogeneous
Euclidean evolution. In those variables the twist that is tuned to impose the
charge constraint is precisely the residual twist $\Delta\eta $, rather than the
total unrotated quasiperiodicity $\eta_b$.

To clarify the conjugate relation between the twist and the charge, consider the trivial homogeneous case first.
From (\ref{eq:SEhom}) it follows immediately that
\begin{align}
    Q=-\frac{dS_E[\phi_i,\bar\phi_i]}{d\eta_i}\ .
\end{align}
For the case of the bounce a similar relation holds:
the Euler-Lagrange equation from varying 
the bounce exponent with respect to $\eta_b$ gives
\begin{align}
    Q=-\frac{dS_E[\phi_b,\bar\phi_b]}{d\eta_b}\ .
\end{align}
The proof is less trivial now. Using the explicit expression for $S_E$, integrating by parts and using the EoMs for $\phi_b$ and $\bar\phi_b$ one gets
\begin{align}
    Q=-\int d^3x \left(\partial_{\eta_b}\phi_b\ \partial_\tau\bar\phi_b+\partial_{\eta_b}\bar\phi_b\ \partial_\tau\phi_b\right)\Big|_{\tau=-\beta/2}^{\tau=\beta/2}\ .
\label{Qcheck}
\end{align}
To evaluate this expression one needs the asymptotic behaviour of the bounce solutions at $\tau\to\pm\beta/2$ which is
\begin{align}
    \phi_b = \frac{\rho_i}{\sqrt{2}}e^{\omega_i \tau -\Delta\eta /2}\ ,\quad 
    \bar\phi_b = \frac{\rho_i}{\sqrt{2}}e^{-\omega_i \tau +\Delta\eta /2}\ ,\quad  (\tau\to-\beta/2)\ ,
    \label{eq:phi_asymp}
\end{align}
that makes explicit that $\Delta\eta $ corresponds to an offset of Euclidean time in the initial configuration,
and
\begin{align}
    \phi_b = \frac{\rho_i}{\sqrt{2}}e^{\omega_i \tau +\Delta\eta /2}\ ,\quad 
    \bar\phi_b = \frac{\rho_i}{\sqrt{2}}e^{-\omega_i \tau -\Delta\eta /2}\ ,\quad  (\tau\to\beta/2)\ .
    \label{eq:phi_asymp2}
\end{align}
It can be checked that these expressions satisfy the appropriate boundary conditions and reflection symmetry and give the correct total twist $\eta_b$. Plugging them in (\ref{Qcheck}) and using (\ref{eq:eta0_eta_cl_relation}) we recover the total charge $Q=V_3 \omega \rho_i^2$.

Furthermore, the twist parameter $\Delta\eta $ is not free but fixed by the charge-matching condition
\begin{equation}
Q=Q_{\rm bounce}(\Delta\eta ),
\label{eq:charge_matching}
\end{equation}
which determines the saddle value of the projected path integral. As in other fixed-charge semiclassical problems, this implies the Legendre-type relation
\begin{equation}
\frac{dS_{Q,\beta}}{dQ}=\Delta\eta  ,
\label{eq:Legendre_relation}
\end{equation}
which follows immediately from $dS_{E,x}/dQ=(dS_{E,x}/d\eta_x) (d\eta_x/dQ)=-Q (d\eta_x/dQ)$ for $x=b,i$, and provides a useful consistency check when the exponent is computed numerically over a range of charges \cite{Levkov:2017paj}.

Finally, in the neutral limit, $Q\to0,\ \Delta\eta \to0,\ \omega\to 0$,
the twist disappears, the hyperbolic phase becomes trivial, the Euclidean closure reduces to periodicity, and one smoothly recovers the ordinary 
Coleman bounce, which at $T=0$ is $O(4)$ symmetric.
The charged case can be understood as a deformation of this picture. The modulus $\rho$ still plays the role of the object that one would naturally identify as the bounce: it is the part of the field that interpolates between the metastable homogeneous background and the bubble interior, and at $T=0$ it is expected to remain approximately $O(4)$-like, at least for sufficiently small charge. The phase, instead, carries the information about the conserved charge. Accordingly, one expects it to be dominated by the homogeneous rotation,
$\arg \Phi(t,\mathbf x)\sim \omega t$, supplemented by corrections that may depend on space and Euclidean time once the tunnelling solution is constructed. In this sense, the modulus controls the bounce-like profile, while the phase encodes how the fixed charge deforms it.

\subsection{A modified Euclidean action and energy barrier interpretation\label{sec:Ebarrier}}

We now show how the extra $\Delta\eta  Q$ term in the tunnelling exponent (\ref{eq:FQbeta_intro_new}) can be very conveniently absorbed into a 
modified Euclidean action \cite{Lee:1994bza}.
Let us write the bounce contribution to $S_{Q,\beta}$ as
\begin{align}
    S_E^{\rm bounce}+\eta_bQ=\int d^4x \left[\frac12 (\partial_\tau\rho)^2-\frac12 \rho^2(\partial_\tau\alpha)^2+
    \frac12(\nabla\rho)^2-\frac12\rho^2(\nabla\alpha)^2+V(\rho)\right]+\eta_b\int d^3x\ \rho^2\partial_\tau\alpha\ .
\end{align}
Using the fact that $\alpha(\pm\beta/2)=\pm\eta_b/2$, as we have seen above, we can write
\begin{align}
    \eta_bQ =\int d^3x \int d\tau\ \partial_\tau (\alpha \rho^2\partial_\tau\alpha)=\int d^4x\ \rho^2\left[(\partial_\tau\alpha)^2+(\nabla\alpha)^2\right] \ ,
\end{align}
where the last expression is obtained by using the equation of motion for $\alpha$, integration by parts and the fact that $\nabla(\rho^2\nabla \alpha)$ vanishes at spatial infinity.
We end up with 
\begin{align}
    S_E^{\rm bounce}+\eta_bQ=\tilde S^{\rm bounce}_E\ ,
\end{align}
where $\tilde S_E$ is the modified Euclidean action \cite{Lee:1994bza}, defined for a generic configuration as
\begin{align}
\tilde S_E\equiv\int d^4x \left[\frac12 (\partial_\tau\rho)^2+\frac12 \rho^2(\partial_\tau\alpha)^2+
    \frac12(\nabla\rho)^2+\frac12\rho^2(\nabla\alpha)^2+V(\rho)\right]\ .
\end{align}
The effect of $\eta_b Q$ is thus to switch all negative contributions in $S_E^{\rm bounce}$ to positive ones.

The contribution of the homogeneous initial state that is subtracted in the tunnelling exponent is 
\begin{align}
    \beta E_i(Q) = \int d\tau \int d^3x \left[\frac12 \rho_i^2\omega_i^2+V(\rho_i) \right]\ ,
\end{align}
which is nothing but $\tilde S^i_E$. Therefore, we can write
\begin{align}
    S_{Q,\beta} = \Delta \tilde S_E = \tilde S^{\rm bounce}_E - \tilde S_E^i\ .
\end{align}

The tunnelling exponent can be interpreted in WKB manner by thinking of $\tau$ as a parameter that labels different
field configurations traversed during the tunnelling process from $\tau=-\beta/2$ (the initial configuration) to $\tau=0$
(the nucleated bubble configuration) and assigning an energy to them. This gives an energy barrier separating initial and final configurations, and the tunnelling exponent can be reproduced as (four times) the area under such a barrier. This was previously done by Bitar and Chang
\cite{Bitar:1978vx,Bitar:1977wy}
for the neutral case and we extend it here for the charged one. The proof of this is quite simple.
The Euclidean energy integral, written in polar field coordinates reads
\begin{align}
    {\cal E} = \int d^3x \left[-\frac12 (\partial_\tau\rho)^2+\frac12 \rho^2(\partial_\tau\alpha)^2+
    \frac12(\nabla\rho)^2-\frac12\rho^2(\nabla\alpha)^2+V(\rho)\right]\ .
\end{align}
Using the fact that this Euclidean energy is conserved during tunnelling, so that
\begin{align}
    \Delta {\cal E} = {\cal E}^{\rm bounce} - {\cal E}^i = 0\ ,
\end{align}
at all $\tau$, we can rewrite the tunnelling exponent as
\begin{align}
    \Delta \tilde S_E = 4 \int_{-\beta/2}^0 d\tau\ \Delta \tilde E\ , 
    \label{eq:BCcharged}
\end{align}
where the modified energy is
\begin{align}
    \tilde E =\int d^3x\left[\frac12(\nabla\rho)^2+\frac12 \rho^2(\partial_\tau\alpha)^2+V(\rho)\right]\ .
\end{align}
Notice that both at $\tau=-\beta/2$ and $\tau=0$ one has $\partial_\tau\rho=0$ and $\nabla\alpha=0$, so that $\tilde E = E$. Therefore,
the energy barrier  $\Delta \tilde E=0$ at the extremes of the integration interval in (\ref{eq:BCcharged}). This is the generalization
of the result of Bitar and Chang to the case of the decay of a globally charged vacuum.

\section{Explicit example\label{sec:exex}}
\label{sec:example}

We now illustrate how to obtain a fixed-$Q$ bounce solution in a concrete scalar potential. Our example has an 
analytic bounce solution for $Q=0$, which provides a nontrivial benchmark for the Euclidean solver.

The scalar potential is piecewise and taken from \cite{Espinosa:2023osv,Espinosa:2018hue}. 
In the field range $0\le \rho \le m$, with $m$ the typical mass scale, 
the potential supports charged $Q-$ball-type configurations and we use this part of the potential to produce the initial homogeneous charged state.
In particular we are interested in the thin-wall limit that produces $Q$-balls of arbitrarily large radius so that its interior can be treated as a homogeneous charged medium as large as desired.  The shape of the potential producing such $Q$-balls is irrelevant for the subsequent tunnelling towards larger field values.
Nevertheless, for completeness we present the potential we used, taken from (section 6.2 of)~\cite{Espinosa:2023osv}\footnote{With $\phi=\rho$, $\phi_0=\rho_i$, $\phi_a=3/2$ and an overall minus sign due to a different convention.}
\begin{equation}
V(\rho)
=
\frac{3\rho^2}{2}
\left[
{2 \over 3 } \rho-1
+
(\rho-1)^2
\log\!
\frac{(\rho-1)\,\rho_i}{\rho\,(\rho_i-1)}
\right]\ ,
\qquad
0\le \rho \le 1\ .
\label{eq:explicit_potential_branch1}
\end{equation}
Here and from now on we set $m=1$ and all dimensionful quantities have to be understood in the appropriate units of $m$.
The parameter $\rho_i\simeq 1$ denotes the field value in the core of the large $Q-$ball. For numerics we take
$\rho_i=1-\epsilon$ with $\epsilon=10^{-13}$. The radius of the $Q$-ball scales as 
$R \simeq \sqrt{(4/3)\log(1/\epsilon)},$ and diverges in the $\epsilon\to0$ limit. This is the regime relevant for the present work: when the charged object is sufficiently large, its interior is well approximated by a homogeneous charged medium, and the tunnelling problem can be treated locally as nucleation out of that homogeneous background rather than as decay of a finite-size $Q-$ball. 

\begin{figure}[t]
    \centering
    \includegraphics[width=0.7\linewidth]{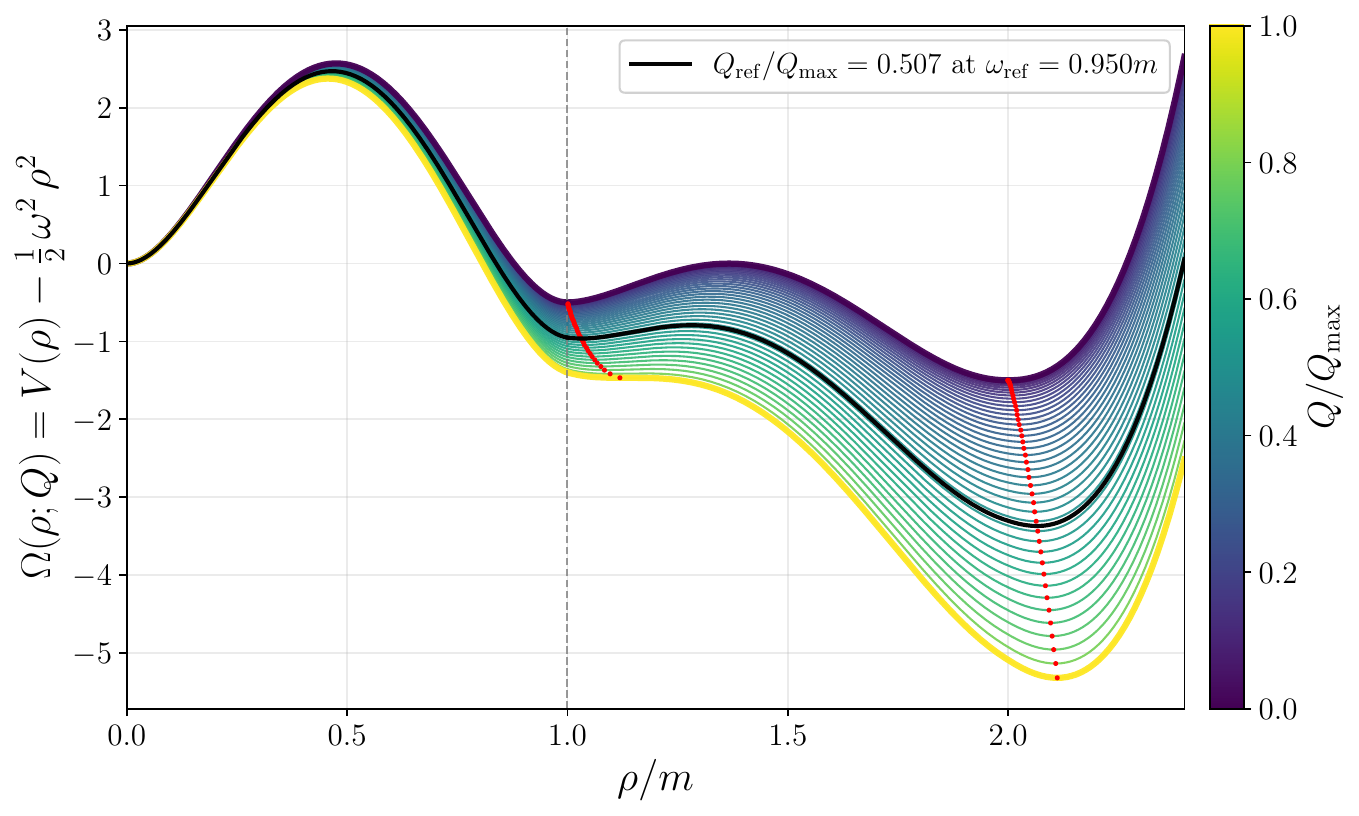}
    \caption{
Reduced homogeneous potential
$\Omega(\rho;Q)=V(\rho)-\omega^2\rho^2/2$
for a family of charged backgrounds with different charges, normalised to $Q/Q_{\rm max}$. Increasing the charge lowers the barrier between the metastable minima near $\rho/m\simeq 1$ and the stable minima near $\rho/m\simeq 2$, (both marked by the red dots) until the false minimum disappears at $Q=Q_{\max}$. The black curve highlights the reference background used in most of the numerical analysis, $Q_{\rm ref}/Q_{\max}=0.507$, corresponding to $\omega_{\rm ref}=0.950m$. 
}
    \label{fig:potential}
\end{figure}

The potential relevant for the decay of such charged medium is
\begin{equation}
V(\rho)
=
-\frac{1}{2}
+
(\rho-1)^2
\left[
2\rho-5
+
(2-\rho)^2
\log\!
\frac{(\rho-2)^2(\rho_0-1)^2}{(\rho-1)^2(\rho_0-2)^2}
\right]\ ,
\qquad
\rho\ge 1\ ,
\label{eq:explicit_potential}
\end{equation}
with $\rho_0<2$. This $V$ has a false vacuum at $\rho=1$ and a true vacuum at $\rho=2$ and we chose it  to have the neutral $Q=0$ decay problem 
under analytic control.\footnote{This $V$ is a version of the potential in Eq.~(29) of \cite{Espinosa:2018hue} with the field shifted by one unit and the energy density 
shifted by $1/2$ for continuity with (\ref{eq:explicit_potential_branch1}) at $\rho=1$.} The parameter $\rho_0$ corresponds to the value of the bounce at its centre, and for numerics we set $\rho_0=1.999$.  The $d=4$ bounce action is also known,
\begin{equation}
S^{(0)}_4
=
-\frac{\pi^2}{3}
\left[
\rho_0-1
+
\operatorname{Li}_2\!\left(
\frac{\rho_0-1}{\rho_0-2}
\right)
\right],
\label{eq:S4_analytic_example}
\end{equation}
and it provides a benchmark for numerics before considering the $Q\neq 0$ cases. 

Before showing the shape of the full potential, we should remember that for $Q\neq 0$, with a nonzero phase-rotation frequency $\omega$, the relevant quantity is the reduced potential 
\begin{equation}
\Omega(\rho;\omega)=V(\rho)-{1 \over 2}\omega^2\rho^2 .
\label{eq:Omega_example}
\end{equation}
In particular this is the combination that enters the EoMs for the $Q$-ball and the charged bounce (at constant $\omega$).
For our choice of potential parameters, $\Omega$ is plotted in Fig.~\ref{fig:potential} for different values of $\omega$ (or equivalently $Q$) 
showing explicitly how finite charge reshapes the tunnelling landscape. The local minima of $\Omega$ define the false and true vacua at fixed $\omega$, $\rho_{\rm false}(\omega)$, $\rho_{\rm true}(\omega)$, which are displaced from their neutral positions. Moreover, as can be seen from the colour coding, the presence of charge lowers the barrier in the effective potential, until the maximal charge is reached, at which point the barrier disappears.
In the following, unless an explicit $\omega$ or $Q$-scan is shown, all plots refer to the reference value $\omega=0.95m$ that corresponds to $Q \simeq 0.507 Q_{\max}$.

The maximal charge $Q_{\max}$, or equivalently, the maximal charge density $q_{\max}=Q_{\max}/V_3$ can be calculated as follows.
For a given
angular velocity $\omega$, one follows the false-vacuum minimum of the reduced
homogeneous potential $\Omega(\rho;\omega)$. As $\omega$ is increased, this minimum eventually disappears by merging with
the barrier. We denote by $\rho_\star$ the value of the radial field at this
spinodal point, and by $\omega_\star$ the corresponding critical phase
velocity. The endpoint is determined by requiring that $\Omega$ have a
stationary inflection point,
\begin{equation}
  {d\Omega\over d\rho}\bigg|_{\rho_\star,\omega_\star}=0,
  \qquad
  {d^2\Omega\over d\rho^2}\bigg|_{\rho_\star,\omega_\star}=0 .
\end{equation}
The maximal charge density is then the homogeneous charge density evaluated at
this endpoint,
\begin{equation}
  q_{\max}=\rho_\star^2\omega_\star .
\end{equation}

For our benchmark potential this can be written explicitly in terms of
$\rho_0$. Defining $\chi_\star \equiv \rho_\star-1$, then the spinodal condition reduces to a single equation for $\chi_\star$,
\begin{equation}
  \left(
    4\chi_\star^3
    +3\chi_\star^2
    -6\chi_\star
    +1
  \right)
  \log\left[
    {(\rho_0-1)(1-\chi_\star)
    \over
    \chi_\star(2-\rho_0)}
  \right]
  +
  4\chi_\star^2
  +5\chi_\star
  -3
  =
  0 .
  \label{eq:spinodal_chi_star}
\end{equation}
Once this equation is solved, $\rho_\star=1+\chi_\star$ and the corresponding
endpoint charge density is
\begin{equation}
  q_{\max}
  =
  (1+\chi_\star)^2
  \left[
    {4\chi_\star(1-\chi_\star)
    \over
    4\chi_\star^3
    +3\chi_\star^2
    -6\chi_\star
    +1}
  \right]^{1/2}.
  \label{eq:qmax_spinodal}
\end{equation}

Note that $\Omega$ is the non-derivative part of the EoM for the modulus $\rho$ for a homogeneous background with given $\omega$.
However, $\Omega$ cannot be considered as a fixed effective potential for the nucleation of the charged bounce because $\omega$ is dynamical: inside the tunnelling bubble, the field approaches a different homogeneous charged state with a different modulus, and charge conservation then requires $\omega$ to change as well, since $q=\omega_i\rho^2_i$ must be conserved.  The role of Fig.~\ref{fig:potential} is thus only to illustrate the reduced potential relevant for the modulus in the initial charged background and it is useful for visualizing how the barrier relevant for nucleation is deformed as the background charge is increased. Naively one might expect that the projection into a fixed-$Q$ sector, as it reduces the available configuration space, must result in a higher tunnelling exponent (in the same way that restricting the domain of a function could lead to higher minima). This is not what happens and, ultimately, the barrier reduction effect just discussed leads instead to a lower exponent.\footnote{However, a conserved charge can also stabilize extended configurations in other contexts. This happens for $Q$-balls, where the fixed charge helps support the lump energetically \cite{Coleman:1985ki,Lee:1991ax}, in false-vacuum-bag-type solitons of the Friedberg--Lee--Sirlin type \cite{Friedberg:1976me} and in cosmological quark-nugget scenarios \cite{Witten:1984rs,Ge:2019voa}.}

Regarding the sign of $Q$, the reduced potential is only sensitive to $\omega^2$, and thus to $Q^2$, so reversing the charge sign does not affect the problem. In particular, the potential barrier is lowered in exactly the same way for positive and negative charge of equal magnitude.\footnote{We thank Diego Redigolo for raising this point in a personal discussion.} Accordingly, all plots involving the charge should be understood as functions of $| Q/Q_{\max}|$.

\subsection{Seeds and solution branches}

The numerical analysis to get the fixed-$Q$ bounce is a saddle-point problem, not a minimization problem, and Newton iteration is highly sensitive to the initial seed. In order to provide good initial seeds it proves useful to study 
first the auxiliary equations
\begin{equation}
\rho''(\xi)+\frac{d-1}{\xi}\rho'(\xi)=\frac{d\Omega}{d\rho}, \qquad \text{for} \quad d= 3,\ 4.
\label{eq:reduced_bounce_eq}
\end{equation}
These are the standard bounce equations  in the effective potential $\Omega(\rho;\omega)$ for a single-variable $\rho(\xi)$ and they play a central organizing role in the full analysis as they correspond to the profiles that the full fixed-$Q$ saddle approaches in different limits. The $d=4$ solution is the most relevant one for the bounce: in the limit $Q\to 0$, it is the $O(4)$ Coleman bounce, and $\rho(\xi)$ provides the natural benchmark for the $\tau=0$ slice of the full $Q\neq 0$ bubble. For small $Q$ the solution deviates only slightly from the $O(4)$ bounce.

The $d=3$ solution is the static critical bubble in the reduced effective potential and is relevant at sufficiently high temperature (or equivalently for sufficiently small $\beta$), when the solution tends to a $\tau$-independent saddle of $O(3)$ type rather than $O(4)$.

A useful way to visualize the relation between these two limits is to vary the Euclidean time size $\beta$. At large $m\beta$, the compact Euclidean direction is effectively infinite on the scale of the bubble size, and the solution is localised well inside the Euclidean box. In this regime the saddle is continuously connected to the zero-temperature $O(4)$ bounce. As $\beta$ is reduced, however, the $O(4)$ configuration can no longer fit inside the compact Euclidean-time interval without interacting with its reflected and periodically copied images. The solution then deforms into a finite-temperature, caloron-like saddle: it is neither fully $O(4)$ symmetric nor yet static, but instead interpolates between the localised quantum bounce and the $O(3)$ thermal critical bubble. In the small-$m\beta$ limit the dependence on $\tau$ becomes weak and the solution approaches the static $O(3)$ branch. This interpolation is shown in Fig.~\ref{fig:calorons}, where the same fixed-$Q$ saddle is reconstructed by reflection around the turning slice and copied along the Euclidean-time direction for different values of $m\beta$.

\begin{figure}
    \centering
    \includegraphics[width=\linewidth]{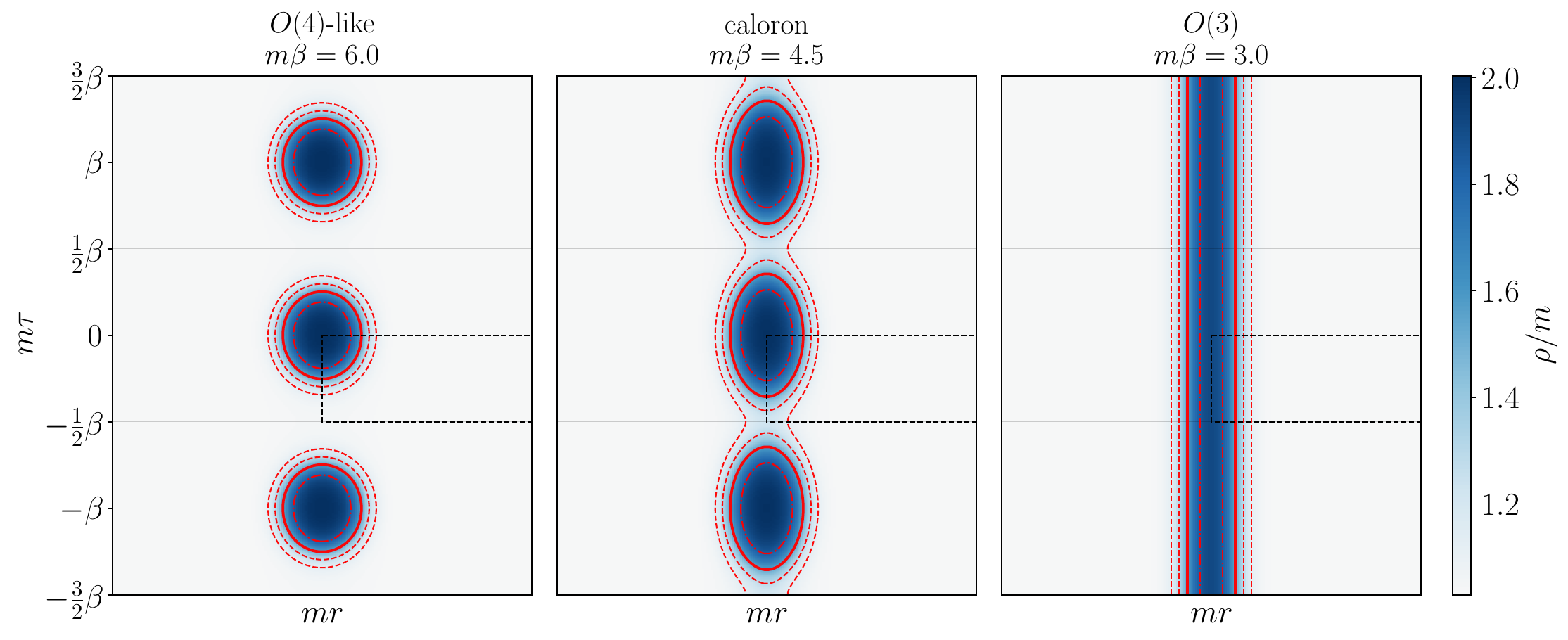}
    \caption{
Evolution of the Euclidean fixed-$Q$ saddle as the temperature is increased from left to right, or equivalently as the temporal size $\beta=1/T$ is reduced.
The colour map shows the modulus $\rho(r,\tau)=\sqrt{2\phi\bar\phi}$, reconstructed by reflection around the turning slice and copied along the Euclidean-time direction.
For large $m\beta$ the solution is localised inside the Euclidean box and is continuously connected to the zero-temperature $O(4)$ bounce.
As $m\beta$ decreases, the $O(4)$ bubble no longer fits inside the compact Euclidean-time interval; the saddle becomes a caloron-like configuration interpolating between the quantum $O(4)$ bounce and the static $O(3)$ thermal critical bubble.
The red contours show constant-$\rho$ levels around the wall, while the dashed black rectangle marks the original half-domain used in the numerical solve.
}
\label{fig:calorons}
\end{figure}

\begin{figure}[t!]
    \centering
    \includegraphics[width=\linewidth]{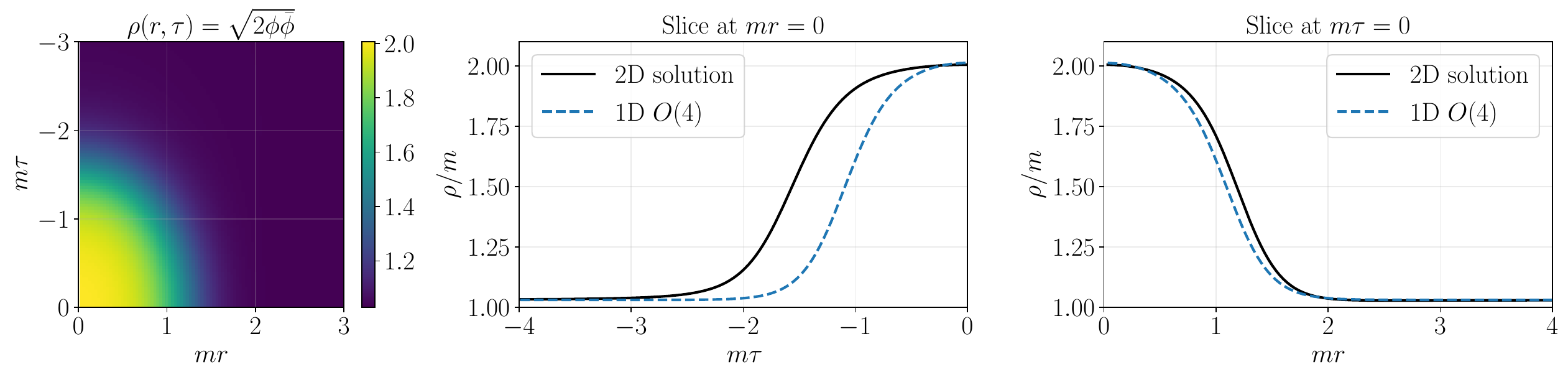}\\
    \includegraphics[width=\linewidth]{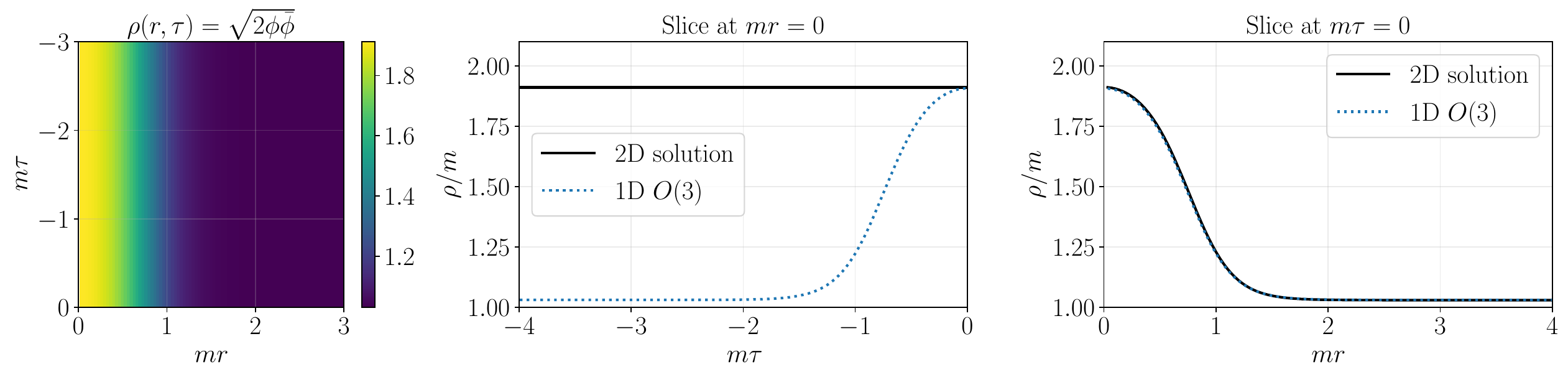}\\
    \caption{We show the final two-dimensional numerical solutions obtained by initializing the solver with different seeds. The left column displays the full converged  solution $\rho(r,\tau)$ on the half-domain, while the last two columns show the corresponding slices at $mr=0$ and $m\tau=0$. In each row, the black solid line denotes the final numerical solution, whereas the coloured lines show the seed/reference profiles used for comparison. \textbf{Top row:} solution obtained from the embedded $O(4)$ seed, corresponding to the branch continuously connected to the neutral Coleman bounce. \textbf{Bottom row:} solution obtained from the $O(3)$ static seed, which converges to a $\tau$-independent configuration. 
    }
    \label{fig:solutions}
\end{figure}

Then, the fixed-$Q$ bounce computation is performed on the half-domain $r\in(0,L_r)$ and $\tau\in\left(-\beta/2,\,0\right)$,
solving the Euclidean EoMs for $\phi$ and $\bar\phi$, Eqs.~\eqref{eq:EOM_phi_O3} and \eqref{eq:EOM_phibar_O3}. One imposes regularity at $r=0$, asymptotic matching to the homogeneous charged background at $r=L_r$, the turning conditions \eqref{eq:turning_conditions} at $\tau=0$, and the twisted-reflected closure condition (\ref{eq:twisted_reflected}) at $\tau=-\beta/2$.

\begin{figure}
    \centering
    \includegraphics[width=0.7\linewidth]{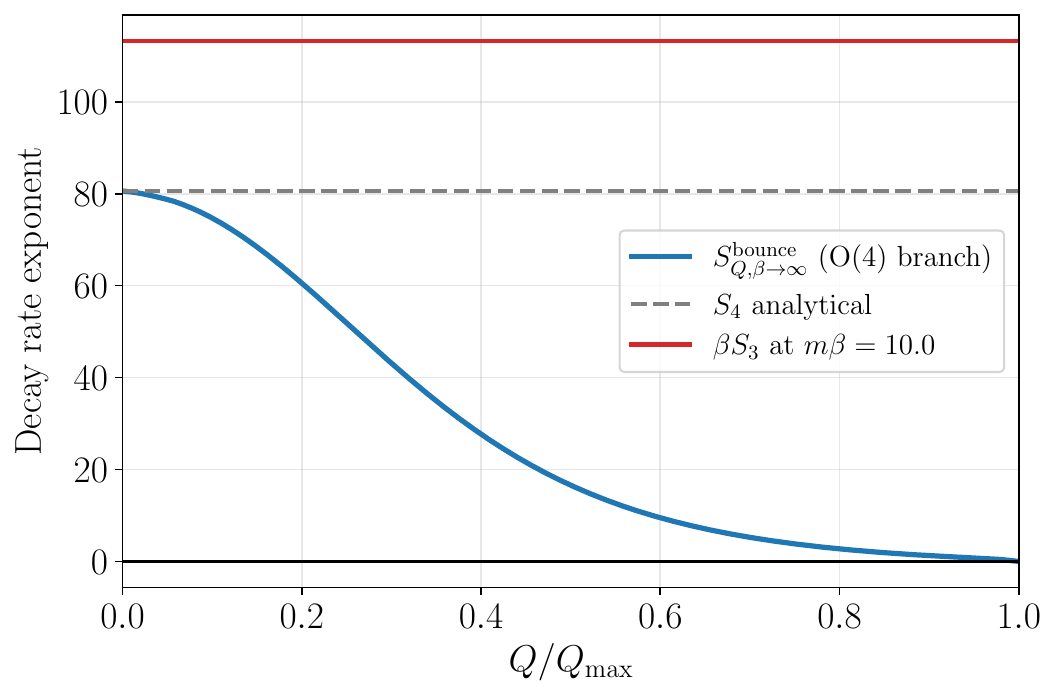}
    \caption{Fixed-$Q$ decay exponent $S_{Q,\beta}$ as a function of the normalised charge $Q/Q_{\max}$, for the $O(4)$ branch used in the explicit analysis. The dashed orange curve is the analytic solution of the quantum bounce for the neutral case in eq. \eqref{eq:S4_analytic_example}, while the red dot-dashed line is the thermal activation in the neutral case evaluated at some low temperature.}
    \label{fig:decay_rate_exponent}
\end{figure}

The logic of the solver is as follows. One first determines the homogeneous charged reference state in the chosen finite box. Then, for a trial value of the twist parameter $\Delta\eta $, one solves the two-dimensional nonlinear PDE problem for $(\phi,\bar\phi)$, computes the charge of the resulting turning slice, and updates $\Delta\eta $ until the target charge is matched.  The $\rho(\xi)$ solutions are used to construct two nontrivial seed families on the $(r,\tau)$ grid. Their role is not just numerical: they probe different nearby branches of the fixed-$Q$ Euclidean EoMs and help to distinguish the charged bounce from other nontrivial solutions of the same PDE problem.

The first seed is the $O(4)$ one, obtained by identifying $\xi$ with the Euclidean radius
$r_E=\sqrt{r^2+\tau^2}$.
This is the natural continuation of the neutral Coleman geometry and therefore the preferred starting point when searching for the charged deformation of the standard bounce.
The second seed is the $O(3)$ static one, obtained by taking the $\rho(\xi)$ $O(3)$ profile and extending it uniformly in $\tau$. A complete description of the ansatz construction is given here \href{https://github.com/GiulioBarni/Qubble}{\faGithub\ Qubble}.

In the explicit example, the solver converges to two qualitatively distinct classes of solutions:
\begin{enumerate}
\item \textit{$O(4)$ branch}: a configuration localised in both $r$ and $\tau$, continuously connected to the neutral bounce, and therefore the natural candidate for the fixed-$Q$ tunnelling saddle. This is shown in the top row of Fig. \ref{fig:solutions}.
\item \textit{$O(3)$ static branch}: an essentially $\tau$-independent solution, shown in the bottom row of Fig.~\ref{fig:solutions}, whose $\tau=0$ slice closely follows the $\rho(\xi)$ $O(3)$ profile. This branch is the charged analogue of a static critical bubble rather than a genuine Euclidean bounce.
\end{enumerate}
A further $O(1)$-like branch is also found
numerically. It is localised mainly in the Euclidean-time direction and is nearly
homogeneous in $r$. Although useful for mapping the nearby branch structure of
the solutions, it does not correspond to a physical bubble-nucleation saddle and
is therefore not reported.

\subsection{Tunnelling action and energetics}
Tunnelling at zero temperature corresponds formally to the limit $\beta\to\infty$. In practice, however, the numerical problem is solved at finite $\beta$, sufficiently large for the solution to approach the zero-temperature regime and for the relevant observables to converge. In this regime, the bounce becomes 
well localised in Euclidean time, its action stabilizes, and the result reproduces the expected $O(4)$ tunnelling configuration, as already shown in Fig.~\ref{fig:calorons}.
The zero-temperature suppression exponent is then $S_Q=\lim_{\beta\to\infty}S_{Q,\beta}$ and the numerical result is given in Fig.~\ref{fig:decay_rate_exponent} by the solid blue line as a function of charge $Q$. As expected, $S_Q$ decreases when the fixed charge is increased, in agreement with the intuition that the charge lowers the barrier between the two vacua, see Fig.~\ref{fig:potential}. For the maximal charge this barrier disappears altogether and $S_Q\to 0$. On the other hand, for $Q\to 0$ we recover the action (\ref{eq:S4_analytic_example}) for the $O(4)$ Coleman bounce (dashed line).

For smaller values of $\beta$, the bounce is stretched in the Euclidean-time direction and may no longer fit properly within the available grid. In that regime, the relevant saddle is no longer the $O(4)$ bounce, but the static $O(3)$ configuration familiar from finite-temperature vacuum decay. The tunnelling exponent is the Boltzmann suppression one, $\beta S_3$, where $S_3$ is the 3-dimensional action of the $O(3)$ bounce. This suppression exponent is given in Fig.~\ref{fig:decay_rate_exponent} for a particular case with $m\beta=10$ by the red line.  The dependence of the tunnelling exponent on $\beta$ is shown in Fig.~\ref{fig:decay_beta}, left plot, with the $O(4)$ branch dominating at large $\beta$.

\begin{figure}
    \centering
    \includegraphics[width=\linewidth]{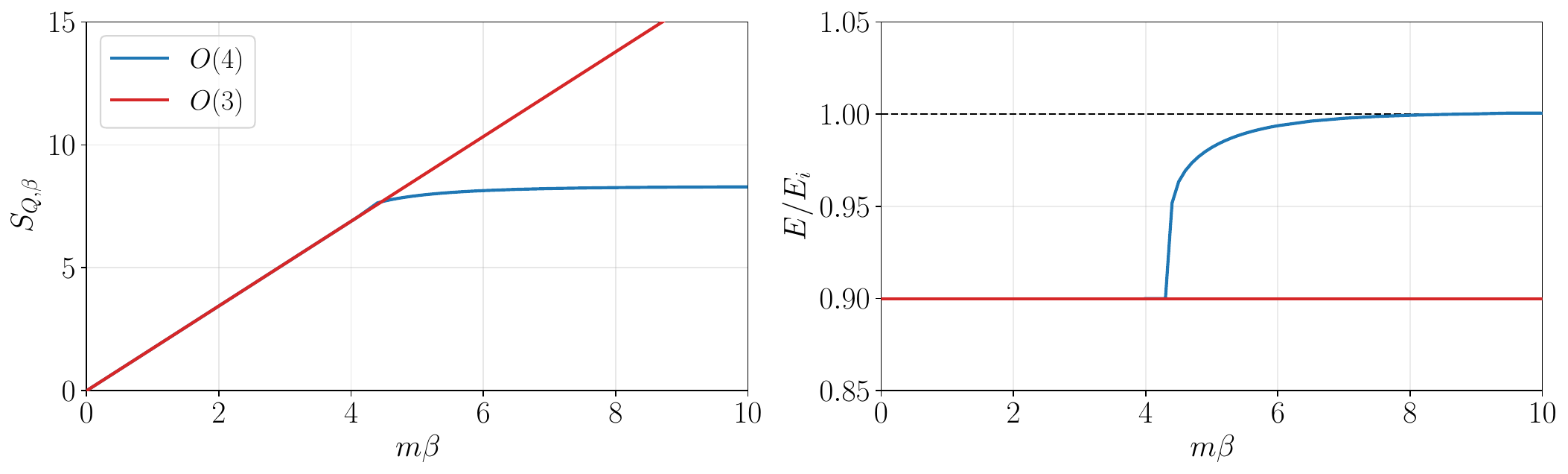}
    \caption{
Dependence of the bounce exponent and energy on the Euclidean time extent $\beta$. 
\textbf{Left}: suppression exponent $S_{Q,\beta}$ extracted from the Euclidean solutions. As $\beta$ grows, the dominant exponent, labelled $O(4)$, converges to a constant value, indicating that the bounce has become insensitive to the finite Euclidean-time interval as the solution approaches the zero-temperature fixed-$Q$ tunnelling configuration. 
\textbf{Right}: energy of the turning slice configuration normalised to the homogeneous energy, showing the approach of the $O(4)$ branch to the homogeneous limit as $\beta$ increases, while the $O(3)$ branch corresponds to the static solution found in Sec.~\ref{sec:example}.
}
    \label{fig:decay_beta}
\end{figure}

Once an Euclidean solution is obtained, one would like to verify explicitly that it has the energetic interpretation expected of a quantum bounce, so that the bubble profile at $\tau=0$, namely the initial profile of the just-nucleated bubble, has exactly the same energy as the initial homogeneous state. At zero temperature, this is precisely what one expects: the bubble is created by quantum tunnelling, not by thermal activation, so there is no thermal bath from which energy to surpass an activation barrier could be borrowed. The critical turning slice must therefore have the same energy as the initial metastable configuration. Fig.~\ref{fig:decay_beta} shows the corresponding energies (normalised to the energy of the homogeneous state) as functions of $\beta$ for the two branches of solutions. In the large-$\beta$ regime, only the $O(4)$ bounce branch approaches the energy conservation expected for zero-temperature decay. By contrast, at smaller $\beta$ the static $O(3)$ branch dominates, consistent with the interpretation in terms of thermal activation rather than zero-temperature tunnelling.

Finally, the reduction of the tunnelling exponent as $Q$ increases can also be understood in terms of the energy of the field configurations under which tunnelling takes place. Following the discussion in Sec.~\ref{sec:Ebarrier}, Fig.~\ref{fig:energy_static} shows the energy barrier in $\Delta \tilde E$ for several values of $Q$. In each case, (four times) the area under each curve corresponds precisely to the tunnelling action, see Eq.~(\ref{eq:BCcharged}), with $\Delta \tilde E(\tau)$ approaching zero at $\tau=-\beta/2$ and again at $\tau=0$, as expected from energy conservation.
The reduction of the $\Delta \tilde E(\tau)$ barrier with increasing $Q$ makes explicit how the conserved charge destabilizes the homogeneous false vacuum.

The turning slice of the Euclidean solution contains the critical bubble with the same conserved energy as the initial homogeneous charged state and is the starting point for the subsequent Minkowski evolution. We will show that finite charge affects this picture in two stages: at small charge, the leading effect is a reduction of the critical bubble radius, while at larger charge, the departure from exact $O(4)$ symmetry becomes an essential part of the solution. The late-time evolution is also important for a second reason: one of the characteristic effects of finite charge is that, even at $T=0$, the bubble does not generically accelerate to the speed of light, but instead approaches a terminal velocity smaller than one. We will see that this follows from the extra energy stored in the charge-carrying phase structure of the wall. To make these points precise, in the next section we will derive a thin-wall estimate for the critical bubble at fixed charge and finally relate the late-time wall energetics to the asymptotic bubble wall velocity.

\section{Thin-wall estimate at fixed charge}
\label{sec:thin_wall_fixed_charge}

In this section we derive a simple thin-wall estimate for the fixed-charge
decay exponent. The aim is to understand the physical origin of the radius
reduction and of the $\Delta\eta  Q$ contribution.

The main point is simple. At fixed charge, a bubble with a larger value of
$\rho$ can store the same charge with a smaller phase velocity. Therefore the
charge energy is reduced inside the bubble. This acts as an additional volume
driving force, reduces the critical radius, and lowers the tunnelling exponent.

\begin{figure}[t!]
    \centering
     \includegraphics[width=0.7\linewidth]{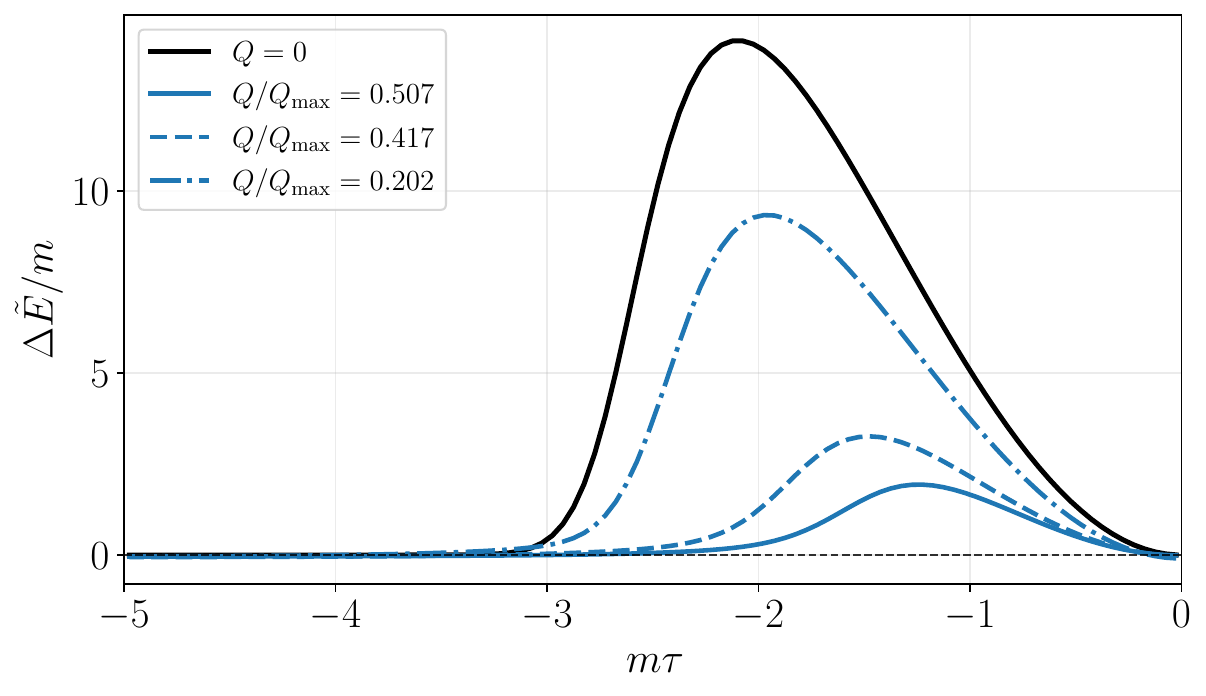}
    \caption{
Modified energy barrier for the $O(4)$-connected charged bounce for different values of $Q/Q_{\max}$, compared with 
the neutral bounce case.  All curves return to zero at the turning slice, as required by energy conservation, while finite charge lowers the
intermediate energy barrier.
}
    \label{fig:energy_static}
\end{figure}

The thin-wall approximation is the regime in which the bubble radius is much larger than the wall thickness. We start with an $O(4)$ symmetric bounce, having in mind the $T=0$ case, with a modulus that can then be approximated by a sharp interface,
\begin{equation}
  \rho(r_E)
  \simeq
  \rho_f\,\Theta(R-r_E)
  +
  \rho_i\,\Theta(r_E-R),
  \qquad
  r_E=\sqrt{r^2+\tau^2},
  \label{eq:rho_step_tw}
\end{equation}
where $i$ denotes the initial phase outside the bubble and $f$ the final
phase inside. The wall contributes through the neutral surface tension
$\sigma_0$, while the neutral bulk driving force is
$\Delta V\equiv V_i-V_f>0$. At $Q=0$, the usual $O(4)$-symmetric
thin-wall action for a radius $R$ is
\begin{equation}
  S_0(R)
  =
  2\pi^2\sigma_0R^3
  -
  {\pi^2\over2}\Delta V R^4\ .
\end{equation}
Imposing $dS_0/dR=0$ we then get the standard results for the bubble radius and action as
\begin{equation}
  R_0={3\sigma_0\over\Delta V},
  \qquad
  S_4^{(0)}
  =
  {27\pi^2\sigma_0^4\over2\Delta V^3}
  =
 {\pi^2\over2}\sigma_0 R_0^3 .
  \label{eq:neutral_tw_action}
\end{equation}
The same result can be obtained from the condition that the energy of the configuration containing the nucleated bubble has the same energy as the initial state.

We will treat the nonzero charge as a small deformation of the neutral bounce, which is the zeroth order of a perturbative expansion. In this
perturbative counting the phase is of order $\epsilon$, while its stress-energy is of order
$\epsilon^2$. Therefore
\begin{equation}
  \alpha
  =
  \epsilon\alpha_1+\mathcal O(\epsilon^3),
  \qquad
  \rho
  =
  \rho_0+\epsilon^2\rho_2+\mathcal O(\epsilon^4).
  \label{eq:small_charge_expansion}
\end{equation}
At order $\epsilon$, there is no ambiguity: one must solve the linear phase
equation on the neutral thin-wall background. At order $\epsilon^2$, the
phase backreacts on the modulus and sources $\rho_2$. As will be shown below, the full function
$\rho_2$ is not needed to compute the on-shell action at this order, because
$\rho_0$ is already a stationary solution. It is, however, necessary to make the
Euclidean energy vanish at generic values of $\tau$. This difference between the action and
the fixed-time energy will be important below. 

\begin{figure}[t]
    \centering
    \includegraphics[width=0.7\linewidth]{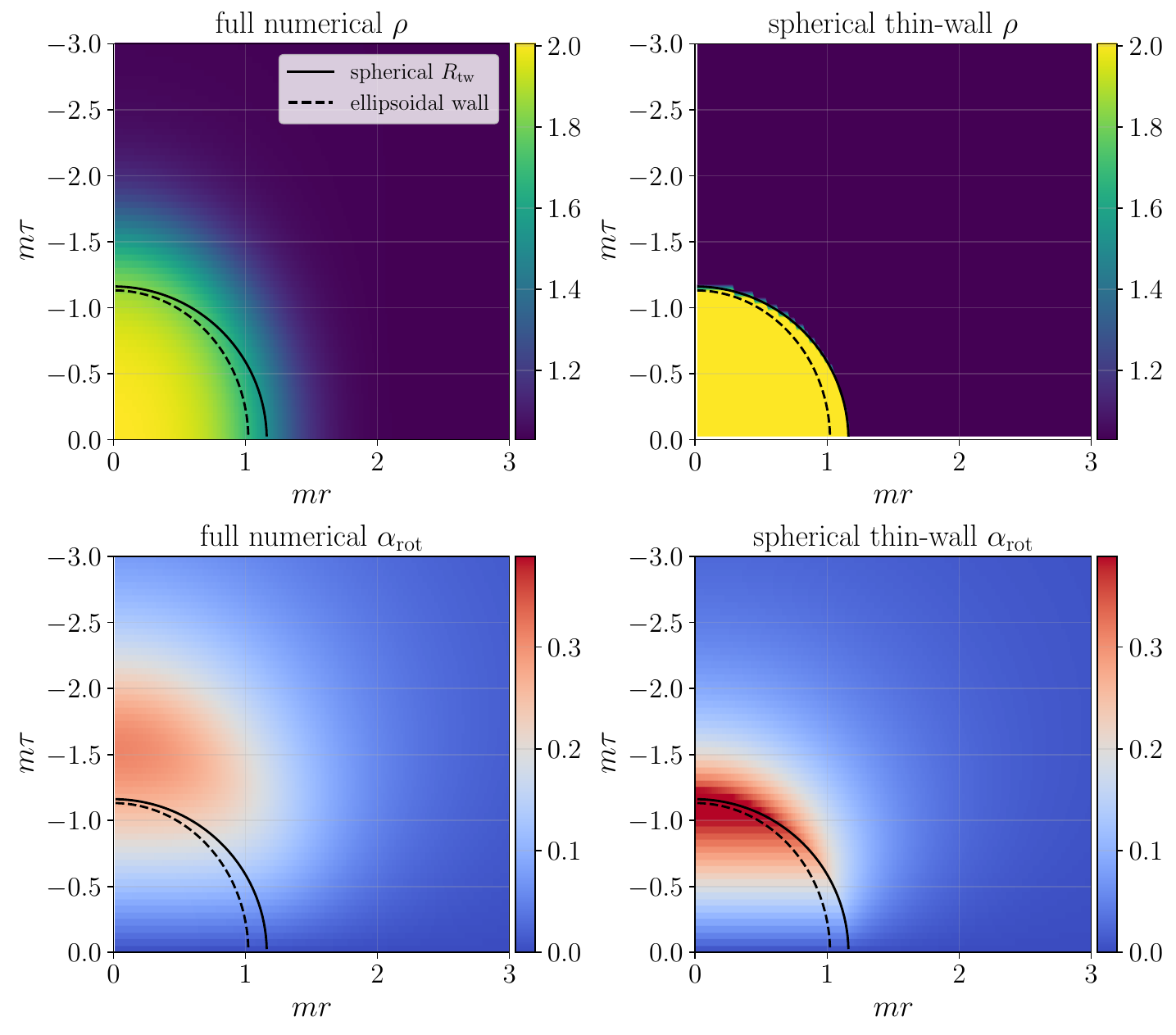}
    \caption{
Full numerical profiles versus the thin-wall approximation.
Top: modulus $\rho$. Bottom: rotated phase
$\alpha_{\rm rot}\equiv\alpha-\omega_i\tau$. The comparison shows how the
step-function for $\rho$ localizes the phase response near the
wall. Solid black line: spherical charged thin-wall radius $R_Q$. Dashed line: collective $O(\epsilon^2)$ modulus backreaction induced by the phase sector, represented as a deformation of the wall into an ellipsoid.
}
    \label{fig:o4_rho_alpha_full_vs_tw_maps}
\end{figure}

\subsection{The Lee phase profile}

As already noticed by Lee in \cite{Lee:1994bza}, even in the thin-wall limit, in which the modulus can be approximated by the sharp profile
\eqref{eq:rho_step_tw}, the phase remains a massless field and its profile, obtained by solving 
\begin{equation}
  \partial_\mu\!\left[
    \rho^2(r_E)\,\partial^\mu\alpha
  \right]=0 \ ,
  \label{eq:phase_dielectric_tw}
\end{equation}
is not a step function. This is the same equation as a dielectric PDE problem in four
Euclidean dimensions, with dielectric coefficient $\rho^2$. The phase and its normal flux, $\rho^2\partial_n\alpha$,
must be continuous at the wall, and these two matching conditions determine the phase
profile.

Although we are interested in the $T=0$ case, it proves convenient to work with
finite $\beta$ and take the $\beta\to\infty$ limit only at the end.
For a bubble much smaller than the Euclidean-time regulator,
$R\ll\beta$, the local phase profile near the wall is insensitive to the
distant boundaries at $\tau=\pm\beta/2$. The only effect of the finite $\beta$ 
is to replace the homogeneous frequency $\omega_i$ by $\omega_b$ given by
\begin{equation}
  \omega_b
  \equiv
  {\eta_b\over\beta}
  =
  \omega_i+{\Delta\eta (R)\over\beta}\ ,
  \label{eq:OmegaR_def}
\end{equation}
in the Lee dielectric solution \cite{Lee:1994bza},
\begin{align}
  \alpha_{\rm out}
  &=
  \omega_b
  \left[
    1
    -
    {\rho_f^2-\rho_i^2\over\rho_f^2+3\rho_i^2}
    {R^4\over r_E^4}
  \right]\tau,
  \qquad
  r_E>R,
  \label{eq:alpha_out_tw}
  \\
  \alpha_{\rm in}
  &=
  \omega_b
  {4\rho_i^2\over\rho_f^2+3\rho_i^2}\,\tau,
  \qquad
  r_E<R ,
  \label{eq:alpha_in_tw}
\end{align}
which satisfies the matching conditions at the wall.
Thus, the phase velocity inside the bubble is fixed by the matching: the bubble polarizes the phase field, and this is the leading thin-wall charge effect. This Lee profile describes how the phase adjusts locally around the bubble, redistributing the charge density between the interior and the exterior. However, this local rearrangement does not determine the total charge of the configuration. The residual twist, $\Delta\eta $, is instead a global quantity that ensures that the bounce configuration has the same total charge as the initial state, see discussion below.

The modulus and phase profiles of both the full numerical
solution and the thin-wall approximation are shown in
Fig.~\ref{fig:o4_rho_alpha_full_vs_tw_maps}. In both cases we plot the rotated
phase $\alpha_{\rm rot}\equiv\alpha-\omega_i\tau$. In the full numerical
solution the deviation from the linear homogeneous behaviour is concentrated in
front of the bubble, on the exterior side of the interface, rather than exactly
on the wall. In the sharp thin-wall approximation, the same response appears
more localised around the wall. This is mostly an artifact of replacing the
smooth numerical modulus by a step function. Solving the same phase equation on
the numerical profile reproduces the displaced structure seen in the full
solution.

\subsection{Fixed charge and the residual twist}

The Lee solution describes the local rearrangement of the phase around a bubble
of given radius. It determines how the charge density is polarised near the
wall, but it does not by itself impose the total charge of the bounce. The
global constraint comes from the fixed-$Q$ projector in the path integral, whose
saddle fixes the total Euclidean twist.

For this reason, we keep the Euclidean-time regulator $\beta$ finite. As discussed in Sec.~\ref{sec:SQbeta}, the
initial homogeneous state has frequency $\omega_i$ and total twist
$\eta_i=\beta\omega_i$, while the bounce has asymptotic frequency $\omega_b$
and total twist $\eta_b=\beta\omega_b$. Their difference,
\[
  \Delta\eta (R)=\eta_b-\eta_i=\beta(\omega_b-\omega_i),
\]
is fixed by requiring that the bounce carries the same total charge as the initial state. Locally, at fixed $\tau$, the difference
$\omega_b-\omega_i=\Delta\eta /\beta$ disappears as $\beta\to\infty$. Globally,
however,  the total accumulated twist $\Delta\eta $ remains finite ($\omega_b$ gives the right asymptotic behaviour $\alpha\simeq \omega_i\tau\pm\Delta\eta /2$ at $\tau=\pm\beta/2$) and contributes to the projected
fixed-charge exponent through the Legendre term $\Delta\eta  Q$. 

Equivalently, the Lee profile may be viewed as the local solution at fixed
asymptotic frequency $\omega_b$, while the saddle condition with respect to the
twist selects the value of $\omega_b$ appropriate to the prescribed charge. On
the turning slice, the charge excess inside the bubble is compensated by the
charge deficit in the exterior Lee tail,
$\Delta Q_{\rm in}+\Delta Q_{\rm out}=0$. Thus the Lee tail changes where the
charge is stored, but the total charge is fixed only after imposing the global
twist condition.

We now compute the charge conjugate to this total twist. For a bubble of radius
$R$, the phase-dependent part of the Euclidean bounce action, has a
homogeneous contribution and a finite Lee polarization contribution,
\begin{equation}
  S_{E,\alpha}^{\rm b}(R,\omega_b)
  =
  -{1\over2}\beta V_3\rho_i^2\omega_b^2
  -
  {\pi^2\over2}R^4
  {\rho_i^2\omega_b^2(\rho_f^2-\rho_i^2)
  \over\rho_f^2+3\rho_i^2}.
  \label{eq:Salpha_bounce_tw}
\end{equation}
Differentiating with respect to $\eta_b=\beta\omega_b$ gives\footnote{
One can also compute the charge on the turning slice as the 3d integral of
$\rho^2\partial_\tau\alpha$ over a large ball of radius $\Lambda$. The correction to the homogeneous charge is then
proportional to $4\pi R^4/\Lambda$, rather than to $\pi^2R^4/\beta$. Both results for the charge (the calculation from the 4d action and the 3d integral)
coincide if temporal and spatial IR cutoffs are related by $\pi\Lambda=4\beta$.
}
\begin{equation}
  Q_{\rm b}
  =-
  {\partial S_{E,\alpha}^{\rm b}
  \over
  \partial\eta_b}
  =
  \left[
    V_3\rho_i^2
    +
    {\pi^2R^4\over\beta}
    {\rho_i^2(\rho_f^2-\rho_i^2)
    \over\rho_f^2+3\rho_i^2}
  \right]\omega_b .
  \label{eq:Q_OmegaR}
\end{equation}
The fixed-charge condition is simply
$Q_{\rm b}  =  Q  =  V_3\rho_i^2\omega_i$. This equation fixes $\omega_b$, and therefore fixes the residual twist
$\Delta\eta (R)$. Using $\Delta\eta =\eta_b-\eta_i$, one obtains
\begin{equation}
  \Delta\eta (R)
  =
  -
  {\pi^2R^4(\rho_f^2-\rho_i^2)\omega_i
  \over
  V_3(\rho_f^2+3\rho_i^2)
  +
  {\pi^2R^4\over\beta}(\rho_f^2-\rho_i^2)}\quad
  \xrightarrow{\beta\to\infty}\quad
  -
  {\pi^2R^4(\rho_f^2-\rho_i^2)\omega_i
  \over
  V_3(\rho_f^2+3\rho_i^2)} .
  \label{eq:eta0_large_beta}
\end{equation}

For $Q>0$ and $\rho_f>\rho_i$, the residual twist is negative. This has a
simple physical meaning. The bubble increases the effective inertia available
to store charge, so the same total charge requires a smaller Euclidean phase
velocity than in the homogeneous false vacuum.

\subsection{Finite--$\beta$
 action and the large--$\beta$ limit}

We now evaluate the thin-wall action using the fixed-charge condition derived
above. The neutral surface and volume terms are unchanged, so the only new
contribution comes from the phase sector. The homogeneous-subtracted Euclidean phase action is
\begin{equation}
  \Delta S_{E,\alpha}(R,\omega_b) \equiv S_{E,\alpha}^b-S_{E,\alpha}^i
  =
  -{1\over2}\beta V_3\rho_i^2
  \left(
    \omega_b^2-\omega_i^2
  \right)
  -
  {\pi^2\over2}R^4
  {\rho_i^2\omega_b^2(\rho_f^2-\rho_i^2)
  \over \rho_f^2+3\rho_i^2}.
  \label{eq:DeltaSalpha_raw}
\end{equation}
This is the phase contribution to $\Delta S_E$, but the
fixed-charge exponent also contains the Legendre term $\Delta\eta Q$.
Thus
\begin{equation}
  \Delta S_{\alpha,Q}^{(\beta)}(R)
  =
  \Delta S_{E,\alpha}(R,\omega_b)+\Delta\eta Q .
  \label{eq:DeltaSalpha_fixedQ_def}
\end{equation}
Using Eq.~\eqref{eq:Q_OmegaR} and \eqref{eq:eta0_large_beta}, the two pieces in \eqref{eq:DeltaSalpha_raw} flip sign 
(as expected on general grounds, see
Sec.~\ref{sec:Ebarrier}) and we finally get
\begin{equation}
  \Delta S_{\alpha,Q}^{(\beta)}(R)
  =
  -
  {Q^2\pi^2R^4(\rho_f^2-\rho_i^2)
  \over
  2V_3
  \left[
    V_3\rho_i^2(\rho_f^2+3\rho_i^2)
    +
    {\pi^2R^4\over\beta}\rho_i^2(\rho_f^2-\rho_i^2)
  \right]} .
  \label{eq:DeltaSalpha_fixedQ_explicit}
\end{equation}
Taking $\beta\to\infty$ and using $Q=V_3\rho_i^2\omega_i$, one obtains
\begin{equation}
  \Delta S_{\alpha,Q}(R)
  =
  -{\pi^2\over2}R^4\delta_Q,
  \qquad
  \delta_Q
  =
  {\rho_i^2\omega_i^2(\rho_f^2-\rho_i^2)
  \over \rho_f^2+3\rho_i^2}.
  \label{eq:deltaQ_large_beta}
\end{equation}
The sign is the expected one: at fixed charge, the converted region stores the
same charge with a smaller phase velocity, and therefore lowers the tunnelling
exponent.

Adding the neutral thin-wall contribution gives
\begin{equation}
  S_{Q,\rm tw}(R)
  =
  2\pi^2\sigma_0 R^3
  -
  {\pi^2\over2}R^4
  \left(
    \Delta V+\delta_Q
  \right).
  \label{eq:SQ_tw_R}
\end{equation}
The stationary radius is
\begin{equation}
  R_Q
  =
  {3\sigma_0\over \Delta V+\delta_Q}
  =
  {R_0\over1+\delta_Q/\Delta V}\ .
  \label{eq:RQ_deltaQ_tw}
\end{equation}
As $\delta_Q>0$, given that $\rho_f>\rho_i$, we see that the radius of the bubble gets reduced by the nonzero charge.

The derivation above is controlled through order $\epsilon^2$, which is the first nontrivial order in the small-charge
expansion. Therefore the strictly perturbative statement would be
\begin{equation}
  R_Q
  =
  R_0
  \left[
    1-{\delta_Q\over\Delta V}
    +O(\epsilon^4)
  \right].
  \label{eq:RQ_smallQ_expansion}
\end{equation}
However, once the leading charge-induced correction to the driving force has been identified, it is natural to keep the stationary thin-wall radius in the unexpanded form of Eq.~\eqref{eq:RQ_deltaQ_tw}. This should be understood as a thin-wall improvement of the collective coordinate $R$, not as a claim that all higher-order corrections in $\omega_i$ have been computed. The reason this improvement is useful is that the radius is determined by a balance between surface and volume terms. Even a leading correction to the volume driving force can produce a sizeable shift of the stationary radius when the charge is moderately large. Keeping Eq.~\eqref{eq:RQ_deltaQ_tw} unexpanded therefore captures the main geometric effect of the charge: the bubble becomes smaller as the effective driving force increases. The full high-charge solution will also contain finite-wall corrections, corrections to the phase profile, and genuine backreaction effects that are not included in the analytic thin-wall estimate.

The validity of this unexpanded thin-wall expression is therefore checked a posteriori against the numerical bounce. As shown in Fig.~\ref{fig:o4_exponent_thin_vs_thick_all_TW}, the unexpanded estimate follows the numerical trend well beyond the very small-charge regime, and the agreement improves as the potential is taken closer to the genuine thin-wall limit. This indicates that the dominant effect over this range is the shift of the effective driving force and of the stationary radius, while the omitted finite-wall and shape corrections appear as subleading deviations.

\subsection{Fixed-charge decay rate}

We can now express the fixed-charge tunnelling exponent directly in terms of the physical
charge ratio
\begin{equation}
  x={Q\over Q_{\max}},
  \label{eq:x_def_tw}
\end{equation}
where $Q_{\max}$ is the endpoint of the homogeneous charged branch, namely the charge at which the metastable barrier disappears, see Sec.~\ref{sec:exex}. The quantity $\delta_Q$ in \eqref{eq:deltaQ_large_beta} grows quadratically with the charge, and we define
\begin{equation}
  {\delta_Q\over\Delta V}
  =
  B x^2 ,
  \label{eq:deltaQ_Bdiel_tw}
\end{equation}
where
\begin{equation}
  B
  =
  {q_{\max}^2\over\Delta V}
  {\rho_f^2-\rho_i^2
  \over
  \rho_i^2(\rho_f^2+3\rho_i^2)},
  \qquad
  q_{\max}={Q_{\max}\over V_3} .
  \label{eq:Bdiel_rhoi_rhof_tw}
\end{equation}
The normalization charge density $q_{\max}$ is determined by the endpoint of the homogeneous charged branch, as detailed in Sec.~\ref{sec:exex}. 
The critical radius is therefore
\begin{equation}
  R_Q
  =
  {R_0\over1+B x^2}.
  \label{eq:RQ_Bdiel_tw}
\end{equation}
Evaluating the fixed-charge action at this radius gives
\begin{equation}
S_{Q,\rm tw}=\frac{\pi^2}{2}\sigma_0 R_Q^3
  \simeq
  {S_4^{(0)}\over
  \left(1+B x^2\right)^3}.
  \label{eq:FQ_Qmax_tw}
\end{equation}
Thus the fixed-charge decay rate takes the closed thin-wall form
\begin{equation}
  {\Gamma_Q\over V}
  \sim
  A_Q
  \exp\left[
    -
    {S_4^{(0)}\over
    \left(1+Bx^2\right)^3}
  \right] .
  \label{eq:Gamma_Q_tw}
\end{equation}

We show the tunnelling exponent as a function of the charge in Fig.~\ref{fig:o4_exponent_thin_vs_thick_all_TW}. For comparison with the numerical decomposition, we also display the
homogeneous-subtracted action, the residual twist term, and their sum. In the
thin-wall estimate, these are
\begin{align}
  S_E^{\rm bounce}-S_E^i
  &\simeq
  S_4^{(0)}
  {1+7B x^2
  \over
  \left(1+B x^2\right)^4},\qquad   
  \Delta\eta Q
  \simeq
  -6S_4^{(0)}
  {B x^2
  \over
  \left(1+B x^2\right)^4},
  \label{eq:eta0Q_Qmax_tw}
\end{align}
Their sum gives \eqref{eq:FQ_Qmax_tw}. The full exponent is the physical
quantity. The separated pieces are useful because the numerical computation
naturally gives both $S_E^{\rm bounce}-S_E^i$ and $\Delta\eta Q$.

\begin{figure}[t]
    \centering
    \includegraphics[width=\linewidth]{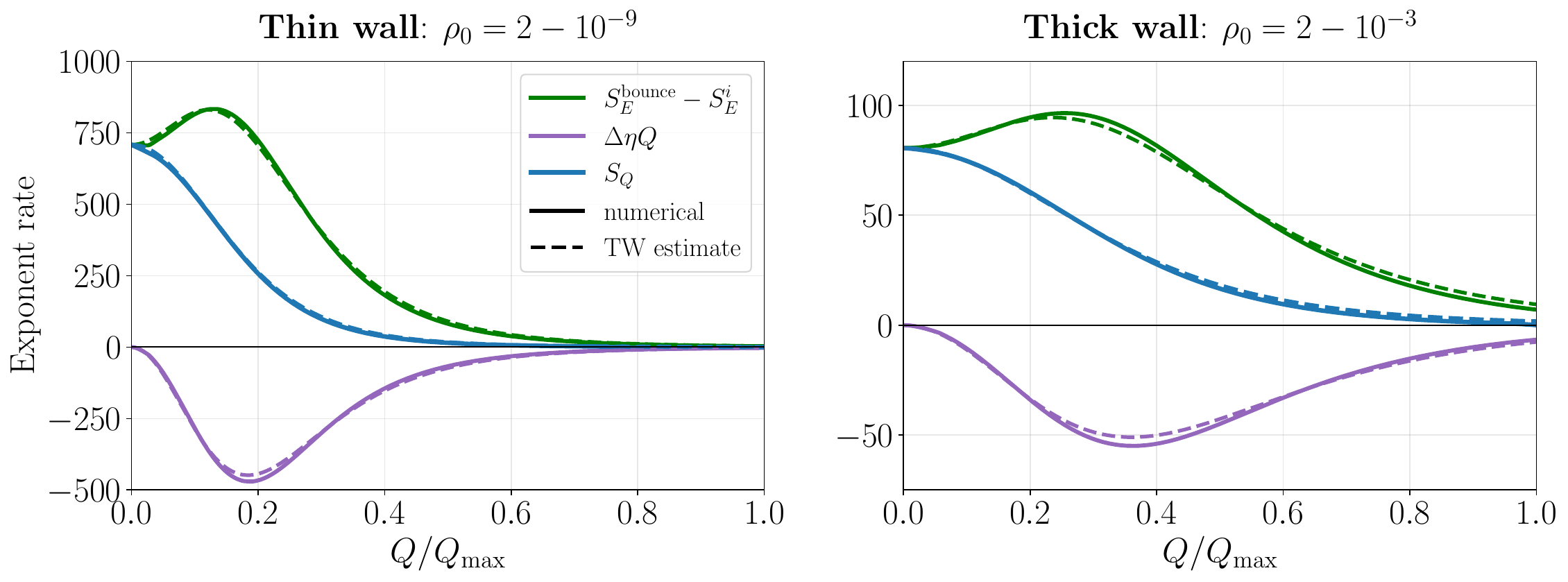}
    \caption{
Thin-wall decomposition of the fixed-charge exponent versus $Q/Q_{\max}$.
The plots show $S_E^{\rm bounce}-S_E^i$, $\Delta\eta Q$, and
$F_Q=S_E^{\rm bounce}-S_E^i+\Delta\eta Q$. Dashed lines are the analytic
thin-wall estimates with the charge coefficient obtained from the Lee
dielectric solution. Left: $\rho_0=2-10^{-9}$, close to the genuine
thin-wall limit. Right: $\rho_0=2-10^{-3}$, the thicker-wall benchmark used
in the rest of the paper.
}
    \label{fig:o4_exponent_thin_vs_thick_all_TW}
\end{figure}

Despite the approximations, the full numerical solutions agree remarkably well
with the thin-wall estimates. This is nontrivial, because the analytic
treatment keeps the phase profile on a fixed spherical $O(4)$ wall and only
represents the modulus response through its collective effect on the energy.
The agreement becomes significantly better when we move closer to the genuine
thin-wall regime of the same class of potentials, as shown in the left panel of
Fig.~\ref{fig:o4_exponent_thin_vs_thick_all_TW}. For the thicker-wall benchmark
used in the rest of the paper, the thin-wall formula still captures the
qualitative trend, although finite-wall and $O(4)$-breaking corrections are
visible.

\begin{figure}[t]
    \centering
    \includegraphics[width=\linewidth]{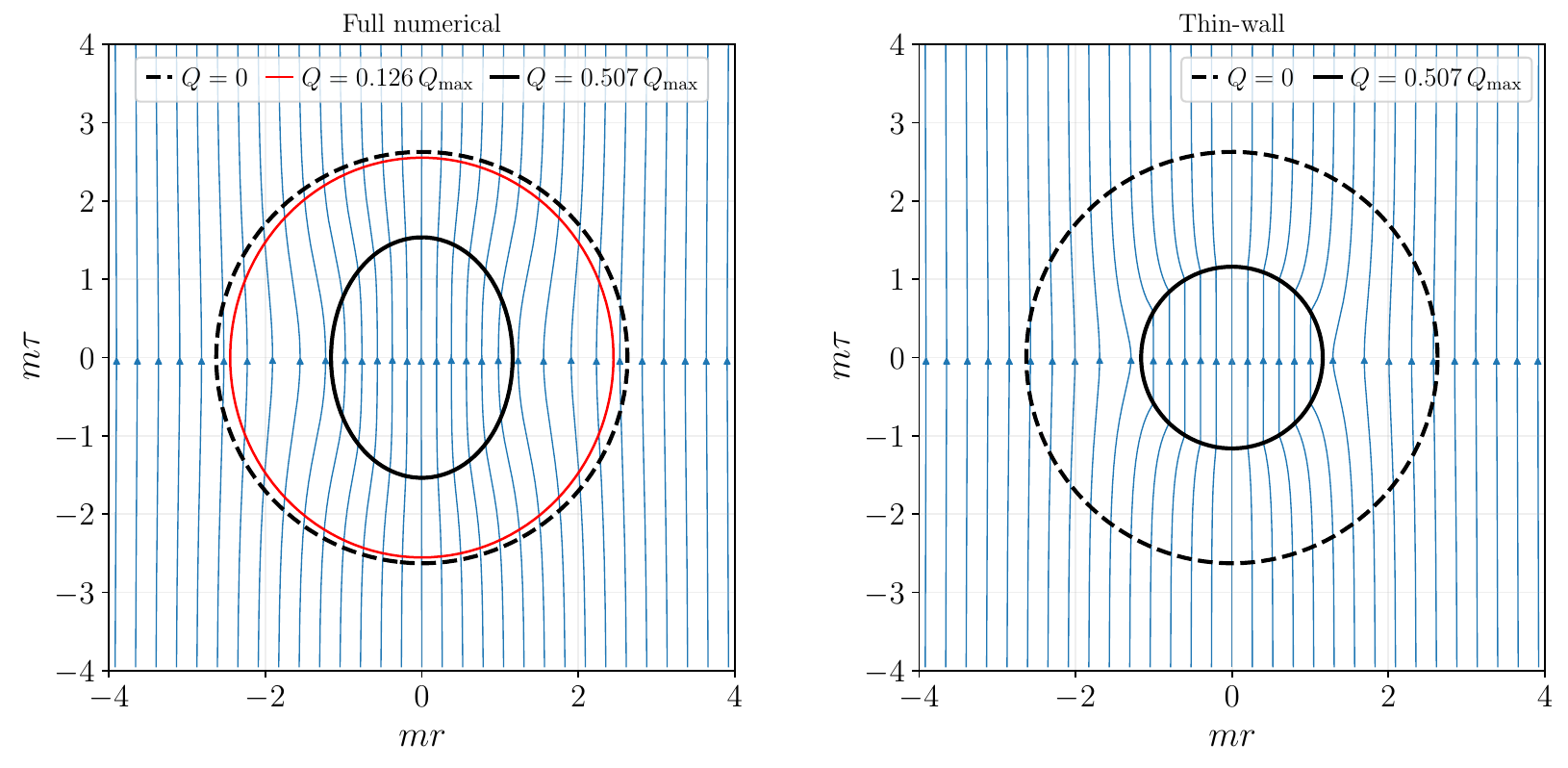}
    \caption{
    Interface contours and Euclidean charge-flow lines for the charged bounce.
    Left: full numerical solutions, compared with the neutral $Q=0$ profile.
    Right: sharp thin-wall dielectric solution at the same reference charge.
    The thin-wall panel shows that smooth phase gradients and charge-flow
    polarization are already present before including finite-wall and
    $O(4)$-breaking effects.
    }
    \label{fig:o4_interface_chargeflow}
\end{figure}

This is illustrated in Fig.~\ref{fig:o4_interface_chargeflow}. The left panel
shows the full numerical interfaces: the charged bounce first shrinks relative
to the neutral Coleman profile and then becomes visibly deformed as the charge
is increased. The right panel shows the sharp thin-wall
dielectric solution. Even with a perfectly spherical wall, the phase equation
already produces smooth charge-flow lines around the bubble. The comparison
therefore separates two effects: the polarization of the phase, already present
in the thin-wall limit, and the additional finite-wall backreaction that
deforms the full numerical bounce.

The blue lines are streamlines of the Euclidean charge current in the
$(r,\tau)$ plane. In the numerical panel they are obtained from
\begin{equation}
  j_\tau
  =
  \bar\phi\,\partial_\tau\phi
  -
  \phi\,\partial_\tau\bar\phi,
  \qquad
  j_r
  =
  \bar\phi\,\partial_r\phi
  -
  \phi\,\partial_r\bar\phi,
\end{equation}
after reconstructing the solution by reflection. In the thin-wall panel they
are computed from the analytic dielectric phase solution
\eqref{eq:alpha_out_tw}--\eqref{eq:alpha_in_tw}. Their role is diagnostic: they
show how the conserved charge is redistributed around the bubble wall.

\subsection{Energy conservation and the role of the modulus response\label{sec:ellipsoid}}

The Euclidean action can be calculated to order $\epsilon^2$ without knowing
the correction $\rho_2$ to the modulus profile because $\rho_0$ is a stationary solution. 
However, the Euclidean energy on a fixed $\tau$-slice is different since it is sensitive to $\rho_2$, 
and the Lee phase alone does not give the full $\epsilon^2$ contribution to the conserved energy.

For a spherical thin-wall bubble of radius $R$, the spatial radius on
a fixed $\tau$ slice is
\begin{equation}
  R_\tau(R)=\sqrt{R^2-\tau^2}.
  \label{eq:a_tau_R_tw}
\end{equation}
The neutral thin-wall energy difference (between bounce and initial state) is
\begin{equation}
  \Delta E_0(\tau;R)
  =
  {4\pi\over3}R^3_\tau(R)
  \left(
    {3\sigma_0\over R}
    -
    \Delta V
  \right).
  \label{eq:DeltaE0_tau_R_tw}
\end{equation}
The Lee phase contribution on the same slice has the form
\begin{equation}
  \Delta E_\alpha(\tau;R)
  =
  -
  {4\pi\over3}R^3_\tau(R)\,
  \delta_E\!\left({\tau\over R}\right)\ ,
  \label{eq:DeltaEalpha_tau_R_tw}
\end{equation}
with
\begin{equation}
    \delta_E=\frac{8\rho_i^2\omega_i^2(\rho_f^2-\rho_i^2)}{5(\rho_f^2+3\rho_i^2)^2}
    \left[2\rho_f^2+3\rho_i^2+3(\rho_f^2-\rho_i^2)\frac{\tau^2}{R^2}\right]\ .
\end{equation}
The fact that  $\delta_E$ depends on $\tau/R$  makes it impossible to satisfy energy conservation
in the form  $\Delta E(\tau;R)=\Delta E_0(\tau;R)+\Delta E_\alpha(\tau;R)=0$ for a $\tau$-independent
bubble radius $R$. The missing term in $\Delta E(\tau;R)$ is precisely the order-$\epsilon^2$
one coming from $\rho_2$. We do not
need the full local function $\rho_2(r,\tau)$. Energy conservation fixes what its collective contribution must do: it must remove the slice dependence of $\delta_E(\tau/R)$ and replace it with the same zero-mode shift $\delta_Q$ that appears in the action. Therefore the complete thin-wall energy should have the form
\begin{equation}
  \Delta E(\tau;R)
  =
  {4\pi\over3}R_\tau^3(R)
  \left(
    {3\sigma_0\over R}
    -
    \Delta V
    -
    \delta_Q
  \right).
  \label{eq:DeltaEfull_tau_R_tw}
\end{equation}
Indeed, we get $\Delta E(\tau;R_Q)=0$ for $R=R_Q$ as given by
\eqref{eq:RQ_deltaQ_tw}. Thus the radius obtained by extremising the action coincides with the
radius for Euclidean energy conservation slice by slice, as it should.

A useful way to understand the missing contribution is to picture it as coming 
from the wall deforming slightly away from a sphere. This deformation must be of order
$\epsilon^2$ and it is part of the modulus response $\rho_2$. We parametrize this collective deformation by
\begin{equation}
  {r^2\over a^2}
  +
  {\tau^2\over b^2}
  =
  1,
  \qquad
  a=R_Q(1+u),
  \qquad
  b=R_Q(1+v),
  \qquad
  u,v=\mathcal O(\epsilon^2).
  \label{eq:ellipsoid_uv_tw}
\end{equation}
At this order, the Lee phase energy can still be evaluated on the spherical
background: the phase energy and the wall deformation are both of order
$\epsilon^2$ and so their product would only contribute at order $\epsilon^4$.

The Lee phase energy contains only a constant term and a term proportional to $\tau^2$. The two parameters
$u$ and $v$ introduce the right $\tau$ dependence and are enough to cancel both. With
\begin{equation}
    a=R_Q\left[1-\frac{6\rho_i^2\omega_i^2(\rho_f^2-\rho_i^2)^2}{5\Delta V(\rho_f^2+3\rho_i^2)^2}\right]\ ,\quad 
    b=R_Q\left[1+\frac{18\rho_i^2\omega_i^2(\rho_f^2-\rho_i^2)^2}{5\Delta V(\rho_f^2+3\rho_i^2)^2}\right]\ ,
    \label{eq:ellispe_coefficients}
\end{equation}
the ellipsoidal wall contribution satisfies
\begin{equation}
  \Delta E_{\rho_2}^{\rm wall}(\tau;R_Q)
  =
  {4\pi\over3}R_\tau^3(R_Q)
  \left[
    \delta_E\!\left({\tau\over R_Q}\right)
    -
    \delta_Q
  \right]\ ,
  \label{eq:wall_response_energy_tw}
\end{equation}
so that
\begin{equation}
  \Delta E\equiv \Delta E_0+\Delta E_\alpha+\Delta E_{\rho_2}^{\rm wall}=0\ ,
  \qquad
  \forall \tau ,
  \label{eq:energy_cancel_ellipsoid_tw}
\end{equation}
up to corrections of order $\epsilon^4$. Notice also that $b-a>0$, implying that 
the ellipsoidal wall is elongated along the $\tau$ direction. This can be seen in Fig.~\ref{fig:o4_rho_alpha_full_vs_tw_maps} in the dashed black lines, and also this is in 
agreement with the numerical findings, see Fig.~\ref{fig:o4_interface_chargeflow}.

This construction should not be read as a full solution for $\rho_2$. It only
captures the collective wall response required by energy conservation. The full
$\rho_2$ also contains local deformations near the wall. Nevertheless, the
argument shows that there is no contradiction between the action result and
energy conservation: the Lee phase profile alone is incomplete for the
fixed-time energy, while the complete order-$\epsilon^2$ solution conserves
energy and selects the same radius $R_Q$.

The same small ellipsoidal deformation does not change the integrated
fixed-charge action at this order. Once the global radius $R_Q$ extremises
the reduced action \eqref{eq:SQ_tw_R}, the first variation of the
action with respect to a small shape deformation vanishes. The deformation is
therefore needed to understand the slice-by-slice energy, but does not modify the
leading thin-wall exponent.

\section{Minkowski evolution}
\label{sec:Minkowski}

The Euclidean saddle, analytically continued to real time, gives the Minkowski 
evolution of the nucleated bubble.  In the neutral case this step is standard: the
bounce reaches a turning slice with vanishing Euclidean-time derivative, and the
data at that slice are used as initial conditions for the subsequent Minkowski
evolution \cite{Coleman:1977py,PhysRevD.16.1762}.  At fixed charge, the logic is
similar, but with the two independent fields, $\phi$ and $\bar\phi$, recombined into a complex field
and with
turning conditions modified accordingly \cite{Lee:1994bza,Levkov:2017paj}.

Before turning to the full Minkowski evolution, it is useful to extract the main
physics analytically in the thin-wall limit.  This approximation already shows
the key modifications of the nucleation problem due to a conserved charge: the phase behaves as a
dielectric field on the bubble background, the charge lowers the effective bulk
cost of forming the bubble, the critical radius is reduced, and the twist term
$\Delta\eta Q$ is essential for obtaining the correct fixed-$Q$ exponent.  These
thin-wall estimates also anticipate the real-time picture: at finite charge the
phase cannot remain a trivial homogeneous rotation, but must rearrange across
the wall.

We then study the full Minkowski evolution numerically.  This late-time
evolution is also dramatically affected by the global charge.  In particular, even in vacuum, the charged
bubble does not generically run away to the speed of light.  Instead, the
phase-gradient energy generated during the expansion leads to a growing
effective wall energy and to an asymptotic terminal velocity smaller than one.

\subsection{Late-time wall energy, tension growth, and terminal velocity\label{sec:latetime}}
\label{sec:energy_conservation_and_charged_wall_dynamics}

To discuss the real-time evolution, it is convenient to use polar variables,
\begin{equation}
\Phi(t,r)=\frac{\rho(t,r)}{\sqrt{2}}\,e^{i\theta(t,r)}.
\label{eq:Mink_polar}
\end{equation}
The Minkowski energy density in these variables is 
\begin{align}
    e(r,t)= \underbrace{\frac12(\partial_t\rho)^2}_{e_{\rho,t}}+ 
    \underbrace{\frac12(\partial_r\rho)^2}_{e_{\rho,r}}+
    \underbrace{\frac12\rho^2(\partial_t\theta)^2}_{e_{\theta,t}}+
    \underbrace{\frac12\rho^2(\partial_r\theta)^2}_{e_{\theta,r}}+
    \underbrace{\frac{}{}V(\rho)}_{e_V}\ .
    \label{eq:mink_energy_pieces}
\end{align}
To isolate the wall contribution, we subtract at each time a piecewise bulk
reference energy density built from the inner and outer plateaus of the evolving profile. We then define the wall energy as
\begin{equation}
E_{\rm wall}(t)
=
4\pi\int dr\,r^2\left[e(r,t)-e^{\rm bulk}(r,t)\right],
\label{eq:Ewall_component}
\end{equation}
where the bulk energy density $e^{\rm bulk}(r,t)$
is constructed as a step-function reference profile at the instantaneous bubble radius $R(t)$.  The inner and outer bulk densities are evaluated from the plateau values of $\rho$ and $\dot\theta$ so that
\begin{equation}
    e^{\rm bulk}(r,t)=
\begin{cases}
V(\rho_{\rm in})+\frac12\rho_{\rm in}^2\dot\theta_{\rm in}^2, & r<R(t),\\
V(\rho_{\rm out})+\frac12\rho_{\rm out}^2\dot\theta_{\rm out}^2, & r>R(t).
\end{cases}
\end{equation}
The remaining integral therefore measures the excess energy localised around the interface, including gradients and kinetic distortions associated with the wall.

If $R(t)$ denotes the bubble radius and
\begin{equation}
\gamma(t)=\frac{1}{\sqrt{1-\dot R^2}},
\end{equation}
the wall tensions (energies per unit surface)} in the wall-frame and rest-frame are
\begin{equation}
\sigma_{\rm wall}(R)=\frac{E_{\rm wall}(R)}{4\pi R^2},
\qquad
\sigma_{\rm rest}(R)=\frac{E_{\rm wall}(R)}{4\pi R^2\gamma(R)}.
\label{eq:sigma_defs}
\end{equation}

\begin{figure}
    \centering
    \includegraphics[width=\linewidth]{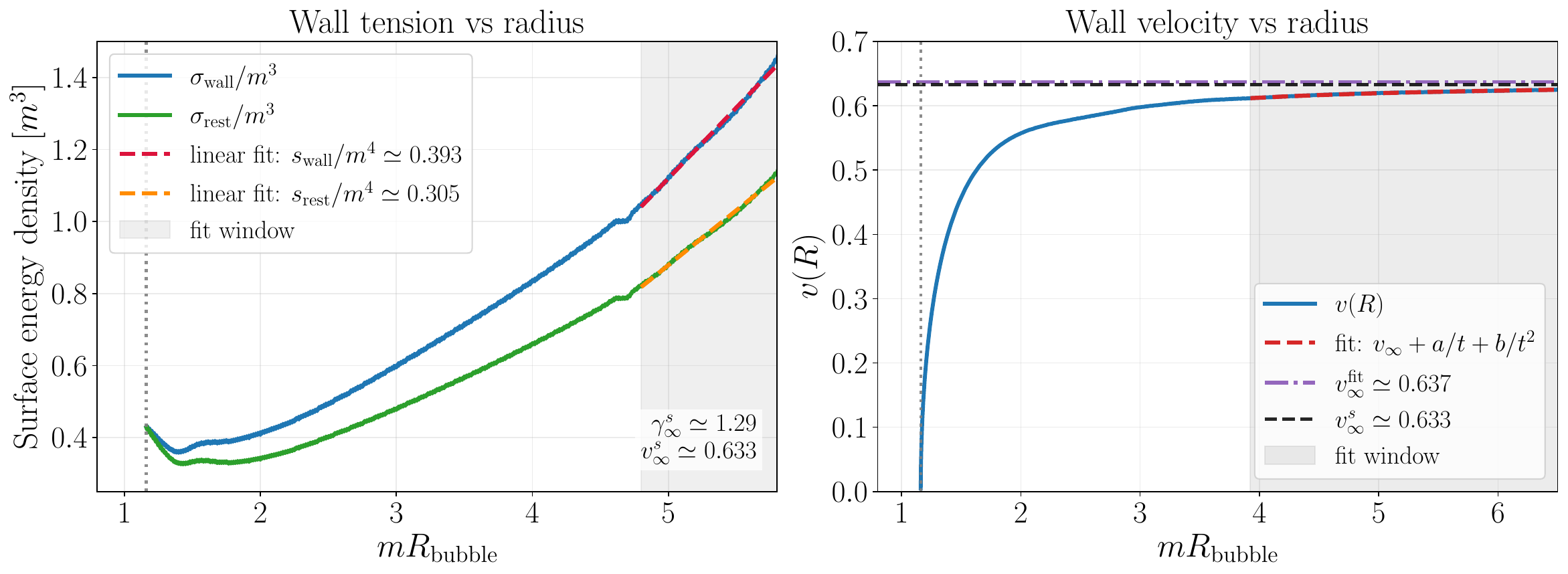}
    \caption{Late-time wall energetics and asymptotic bubble velocity. \textbf{Left}: wall-frame and rest-frame surface energy densities, $\sigma_{\rm wall}(R)$ and $\sigma_{\rm rest}(R)$, plotted as functions of the bubble radius. The key observation is that $\sigma_{\rm rest}(R)$ itself grows linearly. \textbf{Right}: bubble wall velocity as a function of bubble radius, together with the late-time fit used to extract the asymptotic terminal velocity $v_\infty$.}
    \label{fig:asymptotic_velocity_and_surface_tension}
\end{figure}

We find numerically that, after a short transient, the late-time evolution of the wall enters a regime in which the rest-frame wall energy is well fitted by 
\begin{equation}
\sigma_{\rm rest}(R,Q)\simeq \sigma_0(Q)+s(Q)\,R\ ,
\label{eq:sigma_linear}
\end{equation}
as shown in Fig.~\ref{fig:asymptotic_velocity_and_surface_tension}, left plot. In the same figure, right plot,
we show that the same solutions also approach a constant asymptotic wall velocity $v_\infty(Q)<1$. 
These two facts are closely related. At finite charge, the expanding bubble cannot be described only by the usual modulus profile: the phase must also adjust in order to connect two homogeneous states rotating with different angular frequencies.
The phase sector contributes in two physically distinct ways. The temporal phase energy $e_{\theta,t}$
behaves mainly as a bulk rotational-energy term. In an approximately homogeneous region, 
\begin{equation}
q=\rho^2\omega,
\qquad
\omega=\partial_t\theta, \quad \to \quad 
e_{\theta,t}
=
\frac12 q\,\omega
=
\frac{q^2}{2\rho^2}.
\end{equation}
Thus, the same charge can be stored more cheaply energetically in the phase with larger $\rho$ (which plays the same role as the moment of inertia for a rotating body). Since the interior plateau has a smaller angular frequency, $\omega_{\rm in}<\omega_{\rm out}$, the temporal phase term lowers the bulk energy inside the bubble and therefore contributes to the effective pressure difference driving the bubble expansion.

The radial phase-gradient term $e_{\theta,r}$
plays a different role. The numerical profiles show that it remains localised in a thin shell slightly in front of the wall; its support does not broaden macroscopically with $R$. What grows instead is the energy stored in the shell. Numerically, the shell-integrated contribution
\begin{equation}
E_{\theta,r}^{\rm shell}
=
4\pi\int_{\rm shell}dr\,r^2 e_{\theta,r}\ ,
\end{equation}
scales approximately as $R^3$ in the late time evolution. In the effective wall description, this appears as the linear growth of $\sigma_{\rm rest}(R,Q)$ in Eq.~\eqref{eq:sigma_linear}. The growing phase-gradient cost therefore provides an additional wall contribution that competes with the bulk one, driving expansion and is consistent with the observed approach to a terminal velocity, see below.

This also explains why the effect is negligible near nucleation. Close to the turning slice, the bubble is still close to the critical Euclidean configuration, and the phase-gradient shell has not yet built up a large amplitude. The usual surface term therefore dominates. The linear growth of $\sigma_{\rm rest}$ emerges only later, once the expanding wall continuously sources a growing radial phase-gradient cost. A more detailed estimate, together with the numerical diagnostics supporting this interpretation, is given in Appendix~\ref{app:late_time_estimate}.

This late-time behaviour implies a subluminal terminal wall velocity. Neglecting subleading $\mathcal O(R^2)$ terms, the large-$R$ energy of the bubble can be written as
\begin{equation}
\Delta E(R)
\simeq
4\pi R^2\gamma(R)\,\sigma_{\rm rest}(R,Q)
-\frac{4\pi}{3}\Delta P(Q)\,R^3,
\label{eq:DeltaE_largeR}
\end{equation}
where $\Delta P(Q)$ is the pressure difference driving bubble expansion, namely the difference between the inner and outer bulk energy densities evaluated on the late-time plateaus. Using Eq.~\eqref{eq:sigma_linear}, one obtains
\begin{equation}
\Delta E(R)
\simeq
4\pi \gamma(R)\sigma_0(Q)R^2
+
4\pi R^3
\left[
\gamma(R)\,s(Q)-\frac{\Delta P(Q)}{3}
\right].
\label{eq:DeltaE_largeR_expand}
\end{equation}
Since the total energy is conserved, the coefficient of the $R^3$ term must vanish asymptotically. This yields
\begin{equation}
\gamma_\infty(Q)=\frac{\Delta P(Q)}{3\,s(Q)},
\qquad
v_\infty(Q)
=
\sqrt{1-\frac{1}{\gamma_\infty(Q)^2}}
=
\sqrt{1-\left[\frac{3\,s(Q)}{\Delta P(Q)}\right]^2}.
\label{eq:v_inf_Q}
\end{equation}
Thus, the larger the charge-induced slope $s(Q)$ of the wall tension, the smaller the terminal velocity. In the neutral limit, $s(Q)\to0$, one recovers the usual runaway behaviour, $v_\infty\to1$.

\begin{figure}[t!]
\centering 
\includegraphics[width=0.85\linewidth]{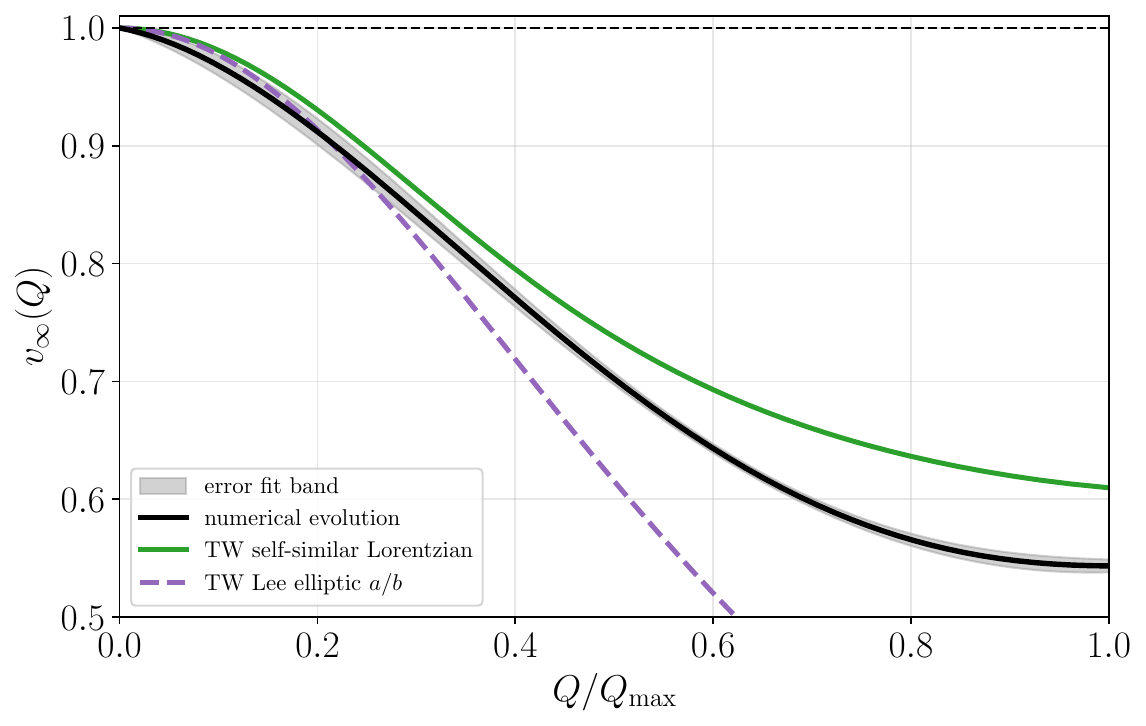} 
\caption{Asymptotic bubble-wall velocity as a function of the charge, normalised to $Q/Q_{\max}$. The black curve shows the central value extracted from the late-time Minkowski evolution, while the shaded band reflects the spread obtained from different velocity estimators and fitting prescriptions. The green curve the Lorentzian self-similar thin-wall estimate, and the dashed purple curve the estimate obtained from the analytic continuation of the Lee ellipsoidal deformation. }
\label{fig:velocity_scan} 
\end{figure}

Finally, for sufficiently small charge one may expand
\begin{equation}
s(Q)=s_1 Q+\mathcal O(Q^2),
\label{eq:alpha_smallQ}
\end{equation}
and, if $\Delta P(Q)$ varies smoothly near $Q=0$, this implies
\begin{equation}
v_\infty(Q)
\simeq
\sqrt{1-\left[\frac{3s_1}{\Delta P(0)}\right]^2Q^2},
\qquad
Q\ \text{small}.
\label{eq:v_inf_smallQ}
\end{equation}
This is consistent with the numerical behaviour displayed in Fig.~\ref{fig:velocity_scan}, where the terminal velocity departs quadratically from unity at small charge.

\subsection{A Lee estimate for the terminal velocity}

The ellipsoidal deformation of the wall discussed in Sec.~\ref{sec:ellipsoid} also suggests a simple analytic estimate for the late-time wall velocity. The idea is to analytically continue the collective
wall surface to Lorentzian time, $\tau\to i t$, as
\begin{equation}
  {r^2\over a^2}+{\tau^2\over b^2}=1\quad \to \quad
  {r^2\over a^2}-{t^2\over b^2}=1\ .
\end{equation}
This gives the wall trajectory
\begin{equation}
  R(t)=a\sqrt{1+{t^2\over b^2}} .
  \label{eq:Lee_wall_trajectory}
\end{equation}
The corresponding velocity is
\begin{equation}
  v(t)=\dot R(t)
  =
  {a t\over b^2\sqrt{1+t^2/b^2}},
\end{equation}
and therefore the asymptotic velocity predicted by this ellipsoidal Lee
continuation is simply
\begin{equation}
  v_\infty^{\rm Lee}
  =
  {a\over b}.
  \label{eq:vLee_ab}
\end{equation}
This reproduces the neutral result $v_\infty=1$ when $a=b$, while an ellipsoid
elongated in the Euclidean-time direction, $b>a$, as the one we find, gives a subluminal terminal
velocity.

Using the semi-axes obtained from the thin-wall energy-conservation condition in Eq.~\eqref{eq:ellispe_coefficients},
one finds
\begin{equation}
  v_\infty^{\rm Lee}
  =
  {1-\epsilon_{\rm ell}\over 1+3\epsilon_{\rm ell}},
  \qquad
  \epsilon_{\rm ell}
  =
  {6\rho_i^2\omega_i^2(\rho_f^2-\rho_i^2)^2
  \over
  5\Delta V(\rho_f^2+3\rho_i^2)^2}.
  \label{eq:vLee_exact_epsilon}
\end{equation}
At small charge, this gives
\begin{equation}
  v_\infty^{\rm Lee}
  =
  1
  -
  {24\rho_i^2\omega_i^2(\rho_f^2-\rho_i^2)^2
  \over
  5\Delta V(\rho_f^2+3\rho_i^2)^2}
  +O(\epsilon^4)\ .
  \label{eq:vLee_smallQ}
\end{equation}

This estimate has a clear physical interpretation. The Lee phase profile makes
the Euclidean critical surface longer in the $\tau$ direction than in the
spatial direction. After analytic continuation, this translates into a
hyperbolic Lorentzian trajectory whose asymptotic slope is smaller than one.
Thus the same charge effect that deforms the Euclidean bubble away from a
sphere also predicts a subluminal wall velocity. 
This velocity estimate is also included in Fig.~\ref{fig:velocity_scan}, which shows it is reasonable at small $Q$.

\subsection{A Lorentzian thin-wall estimate of the terminal velocity}
\label{subsec:lorentzian_tw_terminal_velocity}

The argument in Sec.~\ref{sec:latetime} extracts the terminal velocity from the late-time
growth of the wall energy. We now give a simple analytic estimate of the terminal velocity 
directly in Lorentzian signature.

The thin-wall Lorentzian problem has a simple physical structure. At late times
the wall is approximately self-similar, with constant velocity $v$,
\begin{equation}
  R(t)=vt \ .
  \label{eq:RM_vt_ss}
\end{equation}
The phase profile is affected only inside the
future light cone of the nucleation event. Hence, there are three regions: the
bubble interior, $r<R(t)$; the shell between the wall and the light cone,
$R(t)<r<t$; and the unperturbed exterior, $r>t$. This motivates the
self-similar coordinate $\xi={r/ t}$,
so that the wall is at $\xi=v$, while the causal front is at $\xi=1$.

We solve the phase equation in this moving sharp-wall background. The Lorentzian phase equation is
\begin{equation}
  \partial_t\!\left(\rho^2\,\partial_t\theta\right)
  -
  {1\over r^2}\partial_r\!\left(r^2\rho^2\,\partial_r\theta\right)
  =
  0 .
  \label{eq:lorentzian_phase_eom_ss}
\end{equation}
In the thin-wall approximation, $\rho=\rho_f$ inside the bubble and
$\rho=\rho_i$ outside. The natural late-time ansatz is
\begin{equation}
  \theta(t,r)
  =
  \omega t\,f(\xi),
  \label{eq:selfsimilar_theta_ansatz}
\end{equation}
where $\omega$ is the rotation frequency of the unperturbed false vacuum.
With this ansatz, the phase equation reduces to an ordinary differential
equation for $f(\xi)$. Its solution is constant inside the bubble, while in
the shell it is a homogeneous solution plus a self-similar tail.

The form of the solution is fixed almost completely by the self-similar ansatz.
For constant $\rho$, the Lorentzian phase equation reduces to
\begin{equation}
  \left(1-\xi^2\right)f''+{2\over\xi}f'=0,
  \label{eq:f_selfsimilar_ode}
\end{equation}
whose general solution is
\begin{equation}
  f(\xi)=c_1+c_2\left(\xi+{1\over\xi}\right).
  \label{eq:f_selfsimilar_general}
\end{equation}
We use this solution separately inside and outside the moving wall.

Inside the bubble, $0<\xi<v$, regularity at the origin removes the
$1/\xi$ term. Since the two terms $\xi$ and $1/\xi$ come together in
\eqref{eq:f_selfsimilar_general}, this sets $c_2=0$. Hence the
interior phase velocity is homogeneous,
\begin{equation}
  f_f(\xi)=f_w,
  \qquad
  0<\xi<v .
  \label{eq:f_inside_ss}
\end{equation}
The constant $f_w$ is obtained below from matching at the wall.

In the exterior shell, $v<\xi<1$, the solution must match the unperturbed
rotating false vacuum at the light cone. With our normalization this means
$f(1)=1$. Starting from \eqref{eq:f_selfsimilar_general}, this condition
is implemented by writing the shell solution as
\begin{equation}
  f_i(\xi)
  =
  1+C(v)
  \left(
    \xi+{1\over\xi}-2
  \right),
  \qquad
  v<\xi<1 .
  \label{eq:f_shell_ss}
\end{equation}
The combination in brackets vanishes at $\xi=1$, so the boundary condition at
the light cone is automatic. The coefficient $C(v)$ controls the size of the
phase distortion in the shell.

The first matching condition at the wall is continuity of the phase. This gives
\begin{equation}
  f_w=f_i(v)
  =
  1+C(v)
  \left(
    v+{1\over v}-2
  \right).
  \label{eq:fw_continuity}
\end{equation}
The second matching condition is charge conservation across the moving wall.
Equivalently, the normal component of the charge current must be continuous.
For a wall moving as $r=vt$, the relevant normal derivative is proportional to
$\partial_r+v\partial_t$. Therefore, the condition is
\begin{equation}
  \rho_f^2
  \left(
    \partial_r\theta_f+v\partial_t\theta_f
  \right)
  =
  \rho_i^2
  \left(
    \partial_r\theta_i+v\partial_t\theta_i
  \right), \quad {\rm at}\; \xi=v.
  \label{eq:lorentzian_flux_matching}
\end{equation}
This is the Lorentzian version of the flux-matching condition in the Euclidean
Lee problem. It says that no charge is created or lost on the infinitely thin
interface. After some algebra, this condition fixes the shell amplitude
\begin{equation}
  C(v)
  =
  -
  {(\rho_f^2-\rho_i^2)\,v
  \over
  (\rho_f^2-\rho_i^2)
  (v-1)^2
  +
  \rho_i^2
  {(1-v^2)^2\over v^2}}\ .
  \label{eq:Cv_result_ss}
\end{equation}
This coefficient is the Lorentzian analogue of the Lee polarization
coefficient. It is fixed by two physical requirements: continuity of the phase
and conservation of charge flux across the moving wall. 
\footnote{Although we have just solved the Minkowski version of the Euclidean problem one solves to get the Lee phase profile,
notice that the final result is not obtained by simply
Wick rotating the Euclidean Lee solution. In particular, the latter depends on a constant radius $R$ while for the bubble evolution the radius 
must be a function of time.}

We now compute the energy stored in this phase configuration. The relevant
energy density is
\begin{equation}
  e_\theta
  =
  {1\over2}\rho^2
  \left[
    (\partial_t\theta)^2
    +
    (\partial_r\theta)^2
  \right].
  \label{eq:etheta_lorentzian_ss}
\end{equation}
We subtract the homogeneous false-vacuum phase energy inside the light cone.
Because the solution is self-similar, the remaining energy scales as $t^3$,
\begin{equation}
  E_\theta(t)
  =
  4\pi t^3\,{\cal E}_\theta(v).
  \label{eq:Etheta_selfsimilar_scaling}
\end{equation}
The coefficient ${\cal E}_\theta(v)$ is obtained by inserting
\eqref{eq:f_inside_ss} and \eqref{eq:f_shell_ss} into
\eqref{eq:etheta_lorentzian_ss}. It contains two pieces. The interior piece
measures the change in homogeneous rotational energy inside the converted
region. The shell piece measures the positive radial-gradient cost needed to
connect the interior rotation to the unperturbed exterior. Explicitly,
\begin{equation}
  {\cal E}_\theta(v)
  =
  {\omega^2\over2}
  \left\{
    {v^3\over3}
    \left(
      \rho_f^2 f_w^2-\rho_i^2
    \right)
    +
    \rho_i^2
    \int_v^1 d\xi\,\xi^2
    \left[
      \left(
        f-\xi f'
      \right)^2
      +
      (f')^2
      -
      1
    \right]
  \right\}.
  \label{eq:Etheta_coeff_ss}
\end{equation}
The first term can be negative, because the charge is cheaper to store in the
phase with larger $\rho$. The second term is positive, because the expanding
bubble must build a radial phase profile. The competition between these two
terms gives the net shell energy.

At late times, the released vacuum energy also scales as $t^3$,
\begin{equation}
  E_{\rm vac}(t)
  =
  {4\pi\over3}v^3t^3\,\Delta P,
  \label{eq:Evac_selfsimilar_ss}
\end{equation}
where $\Delta P$ is the late-time bulk driving pressure, defined from the
inner and outer plateaus of the evolving solution. A terminal solution is possible when the leading $t^3$ contribution to the conserved energy cancels.
This gives the Lorentzian thin-wall condition
\begin{equation}
  {\cal E}_\theta(v)
  =
  {v^3\over3}\Delta P .
  \label{eq:selfsimilar_terminal_condition}
\end{equation}
This equation determines the terminal velocity in the self-similar thin-wall
estimate. This velocity is shown in Fig.~\ref{fig:velocity_scan} and gives our best approximation to the numerical result.

It is useful to connect this result with the wall-energy language used above.
Since $R=vt$, Eq.~\eqref{eq:Etheta_selfsimilar_scaling} can be written as
\begin{equation}
  E_\theta(t)
  =
  4\pi R^3\gamma\,s_{\rm TW}^{\rm ss}(v),
  \qquad
  s_{\rm TW}^{\rm ss}(v)
  =
  {{\cal E}_\theta(v)\over v^3\gamma}.
  \label{eq:alphaTWss_def}
\end{equation}
Then the terminal condition becomes
\begin{equation}
  \gamma\,s_{\rm TW}^{\rm ss}(v)
  =
  {\Delta P\over3}.
  \label{eq:terminal_alphaTWss}
\end{equation}
This has the same form as the phenomenological condition
\eqref{eq:v_inf_Q}. The difference is that here the coefficient
$s_{\rm TW}^{\rm ss}(v)$ is not fitted from the numerical wall energy.
It is computed from an explicit Lorentzian self-similar phase profile.

The small-charge limit is immediate. The self-similar deformation of the phase
is proportional to the background rotation frequency, so the shell energy is
quadratic in the charge. Equivalently,
\begin{equation}
  {\cal E}_\theta(v,Q)
  =
  Q^2\,{\cal E}_2(v)+\mathcal O(Q^4),
  \label{eq:Etheta_smallQ_scaling}
\end{equation}
where ${\cal E}_2(v)$ is independent of $Q$ at leading order. In the same
limit the driving pressure approaches its neutral value, $\Delta P(0)$.
Solving \eqref{eq:selfsimilar_terminal_condition} then gives a velocity that approaches the speed of light. In terms of the Lorentz factor,
\begin{equation}
  \gamma_\infty(Q)
  \propto
  {1\over Q},
  \qquad
  Q\to0 .
  \label{eq:gamma_smallQ_scaling}
\end{equation}
Therefore
\begin{equation}
  v_\infty(Q)
  =
  1-\mathcal O(Q^2),
  \label{eq:v_smallQ_scaling}
\end{equation}
which is the same small-$Q$ behaviour observed in the numerical scan of
Fig.~\ref{fig:velocity_scan}.

\subsection{Initial data from the turning slice}
The thin-wall discussion gives the qualitative picture, but the subsequent
real-time evolution must be initialised from the full Euclidean solution.  We now
describe how the numerical bounce is continued to Minkowski time.  The key point
is that the physical initial data are not imposed by hand: they are read directly
from the turning slice of the fixed-charge saddle, where the Euclidean solution
has the correct reflection properties to define a real-time configuration.

The turning slice is located at $\tau=0$ and is characterised by
\begin{equation}
\phi(0,r)=\bar\phi(0,r),
\qquad
\partial_\tau\phi(0,r)=-\partial_\tau\bar\phi(0,r).
\label{eq:Mink_turning_conditions}
\end{equation}
These relations ensure that the Euclidean configuration can be matched to a physical Minkowski field obeying the usual complex-conjugation condition.

For $t>0$ we analytically continue the Euclidean solution according to
$t=-i\tau$, or equivalently $\tau=it$, and define
\begin{equation}
\Phi(t,r)=\phi(it,r),
\qquad
\Phi^\ast(t,r)=\bar\phi(it,r).
\label{eq:Mink_analytic_cont}
\end{equation}
At the initial time, one obtains
\begin{equation}
\Phi(0,r)=\phi(0,r)=\bar\phi(0,r),
\label{eq:Mink_init_field}
\end{equation}
together with
\begin{equation}
\dot\Phi(0,r)=i\,\partial_\tau\phi(0,r),
\qquad
\dot\Phi^\ast(0,r)=i\,\partial_\tau\bar\phi(0,r),
\label{eq:Mink_init_velocity}
\end{equation}
where the dot denotes the Minkowski-time derivative. Because of
\eqref{eq:Mink_turning_conditions}, these data satisfy
\begin{equation}
\Phi^\ast(0,r)=\Phi(0,r)^\ast,
\qquad
\dot\Phi^\ast(0,r)=\dot\Phi(0,r)^\ast,
\end{equation}
so the post-tunnelling evolution is described by a single physical complex field.

By a global $U(1)$ rotation one may choose $\theta(0,r)=0$, so that the turning slice is specified by
\begin{equation}
\rho_\ast(r)\equiv \rho(0,r),
\qquad
\dot\rho(0,r)=0,
\label{eq:turning_modulus}
\end{equation}
while the initial phase velocity remains nonzero and carries the conserved charge,
\begin{equation}
q_M(0,r)=\rho_\ast^2(r)\,\dot\theta(0,r).
\label{eq:turning_charge_density}
\end{equation}
This is the key difference with the neutral case: the turning slice is a turning point only for the modulus. The phase rotation does not stop because it is precisely what stores the fixed charge. In the neutral problem, the turning slice is often described loosely as a ``static bubble''. At fixed charge, this terminology can be misleading. The modulus is momentarily static, but the phase is not: the configuration already contains the time dependence required to support the conserved charge. For this reason, the real-time evolution of the charged bubble differs qualitatively from the neutral one even when the modulus profile looks close to a standard critical bubble at $t=0$.

\subsection{Real-time equations and conserved quantities}

For a spherically symmetric evolution, the Minkowski equation of motion is
\begin{equation}
\ddot\Phi
-
\partial_r^2\Phi
-
\frac{2}{r}\partial_r\Phi
+
V'(|\Phi|^2)\Phi
=
0\ .
\label{eq:Mink_EOM_Phi}
\end{equation}
Equivalently, in polar variables, one has
\begin{equation}
\ddot\rho-\partial_r^2\rho-\frac{2}{r}\partial_r\rho
-\rho\,\dot\theta^{\,2}
+\rho\,(\partial_r\theta)^2
+\frac{dV}{d\rho}
=
0,
\label{eq:Mink_EOM_rho}
\end{equation}
\begin{equation}
\partial_t(\rho^2\dot\theta)
-
\frac{1}{r^2}\partial_r\!\left(r^2\rho^2\partial_r\theta\right)
=
0.
\label{eq:Mink_EOM_theta}
\end{equation}
The second equation is the local charge-conservation law.

\begin{figure}[t!]
    \centering
    \includegraphics[width=\linewidth]{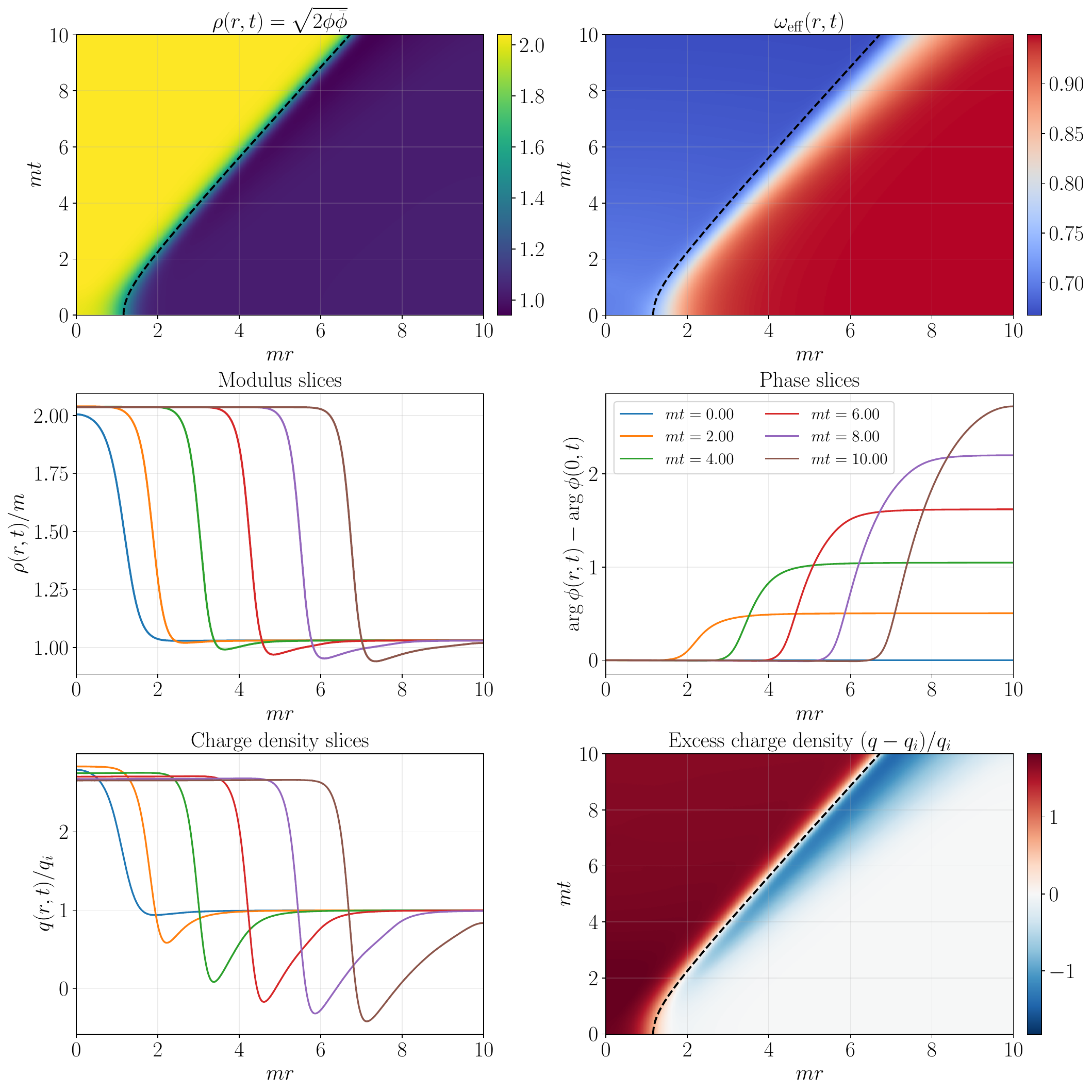}
   \caption{
Post-tunnelling Minkowski evolution of the charged bubble. The upper-left panel shows the modulus $\rho(r,t)=\sqrt{2\phi\bar\phi}$, with the dashed line marking the extracted bubble radius, while the upper-right panel shows the effective local rotation rate $\omega_{\rm eff}(r,t)$. The middle panels show radial snapshots of the modulus and of the phase difference $\arg\phi(r,t)-\arg\phi(0,t)$, after removing the overall rotation at the centre of the bubble. The lower-left panel shows the corresponding charge-density profiles, normalised to the initial homogeneous charge density $q_i$. The lower-right panel displays the redistribution of charge during the evolution: the charge excess inside the bubble is compensated by an opposite charge deficit outside it, while the total charge remains conserved.}
   \label{fig:minkowski_combined_evolution}
\end{figure}

The two conserved quantities that are most useful in practice are the total charge and the Minkowski energy. For the evolution generated from the Euclidean turning slice, these quantities must remain constant up to numerical errors and finite-box effects. 

The analytic continuation of the fixed-$Q$ saddle should be interpreted as the onset of bubble growth in a charged medium. At $t=0$, the modulus profile is stationary, but the phase profile already carries nontrivial charge density and phase current. The subsequent evolution determines whether the configuration is in the basin of attraction of an expanding bubble or instead relaxes back toward the homogeneous branch.

\subsection{Observables in the real-time evolution}

In practice, the most useful observables for the Minkowski evolution are:
\begin{enumerate}
\item the bubble radius $R(t)$, extracted from a fixed level set of the modulus profile;
\item the wall velocity $\dot R(t)$ and Lorentz factor $\gamma(t)$;
\item the conserved charge and energy, used as stability checks;
\item the wall-energy decomposition introduced in the previous section.
\end{enumerate}
Together, these quantities determine whether the continued solution behaves as a genuine charged bubble and whether its late-time dynamics are consistent with the thin-wall picture. This behaviour is illustrated in Fig.~\ref{fig:minkowski_combined_evolution}: the modulus profile shows the outward expansion of the nucleated bubble, while the upper-right panel shows the effective phase-rotation quantity $\omega_{\rm eff}(r,t)=\arg_{\rm unwrap}\phi(r,t)/t$, where the phase has been unwrapped in time to remove the artificial $2\pi$ jumps of the principal branch.\footnote{At each fixed radius, we take the phase $\arg \phi(r,t)$ in $(-\pi,\pi]$ and make it continuous in time: whenever two consecutive times differ by more than $\pi$, we add or subtract $2\pi$. The result is a continuous ``unwrapped'' phase, $\arg_{\rm unwrap}\phi(r,t)$.} Details on the numerical implementation of the Minkowski evolution can be found here \href{https://github.com/GiulioBarni/Qubble}{\faGithub\ Qubble}.

\section{Extension to finite temperature and finite chemical potential}
\label{sec:extension}

The fixed-charge construction discussed above can be extended to thermal and dense environments, as already pointed out in \cite{Levkov:2017paj}, but it is important to separate two logically distinct effects. A finite temperature makes the Euclidean time direction compact, with extent $\beta=1/T$,
and the scalar potential entering the Euclidean equations should be replaced by the appropriate finite-temperature effective potential. Thus $\beta$ should no longer be viewed merely as a regulator used to approach the zero-temperature limit. Rather, each value of $\beta$ corresponds to a physical temperature, and therefore to a distinct thermodynamic state of the system. The fixed-$Q$ saddle is then computed on a finite Euclidean interval, with the same charge projection and twisted boundary conditions as before. In the large-$\beta$ limit, the solution approaches the zero-temperature quantum bounce, whereas for sufficiently small $\beta$ the dominant saddle can become the usual static $O(3)$ thermal configuration, see Fig.~\ref{fig:decay_beta}. In this sense, finite temperature changes the background and the Euclidean geometry of the problem, but not the logic of the fixed-charge projection.

A finite chemical potential requires a more careful distinction. The first possibility is that the chemical potential is associated with degrees of freedom other than the tunnelling field. For instance, the order parameter that nucleates may be neutral, while other species in the medium carry the conserved charge. If those charged species are integrated out, their effect is encoded in a finite-density effective thermal potential,
\begin{equation}
V(\rho)
\;\longrightarrow\;
V_{\rm eff}(\rho;T,\mu).
\label{eq:Veff_T_mu}
\end{equation}
In this case, the bounce of the tunnelling field does not involve a fixed-charge twist of that field. The chemical potential modifies the homogeneous solutions, the barrier, the wall tension, and the nucleation exponent through $V_{\rm eff}$, but one should not impose twisted boundary conditions on a neutral order parameter. If one instead wants to keep fixed the conserved charge carried by the
integrated-out species, rather than its chemical potential, one should first trade the grand-canonical potential for a canonical free-energy density. In
practice, this means computing
\begin{equation}
    n(\rho;T,\mu)
=
-\frac{\partial V_{\rm eff}(\rho;T,\mu)}{\partial \mu},
\end{equation}
inverting this relation to express $\mu$ in terms of $n$, and using the
Legendre-transformed potential
\begin{equation}
V_{\rm eff}^{(n)}(\rho;T,n)
=
V_{\rm eff}(\rho;T,\mu)
+
\mu n    
\end{equation}
evaluated at $\mu=\mu(\rho;T,n)$. For an inhomogeneous bubble, this may
have to be formulated as a constrained variational problem, in which the
density profile $n(r)$ is varied together with the order parameter, while
the total charge $\int d^3x\,n(r)$ is held fixed. We leave a systematic study of the different ways in which chemical potentials and finite density for additional species can be incorporated into the effective tunnelling problem to future work.

The second possibility, which is the one relevant for the complex scalar studied in this work, is that the chemical potential is conjugate to the charge carried by the tunnelling field itself. A homogeneous charged state is then a rotating configuration, as in \eqref{eq: initial_state}.
In a grand-canonical equilibrium with an external reservoir fixing $\mu_{\rm ext}$, the thermodynamic potential is
\begin{equation}
\Omega_{GC}
=
E-\mu_{\rm ext}Q .
\label{eq:grand_potential_general}
\end{equation}
For the homogeneous ansatz \eqref{eq: initial_state}, the grand-potential density is
\begin{equation}
\omega_{\rm GC}(\rho_i,\omega;\mu_{\rm ext})
=
V(\rho_i^2)
+
\frac{1}{2}\omega^2\rho_i^2
-
\mu_{\rm ext}\omega\rho_i^2 .
\label{eq:grand_potential_density_mu}
\end{equation}
Stationarity with respect to $\omega$ gives
\begin{equation}
\frac{\partial \omega_{\rm GC}}{\partial \omega}
=
(\omega-\mu_{\rm ext})\rho_i^2=0,
\end{equation}
and hence, for $\rho_i\neq0$,
\begin{equation}
\omega=\mu_{\rm ext}.
\label{eq:omega_equals_mu_ext}
\end{equation}
Thus, in grand-canonical equilibrium, the chemical potential fixes the rotation frequency of the homogeneous charged state. The charge of the initial state is then not an independent input, but is selected by $\mu_{\rm ext}$
\begin{equation}
Q_i(\mu_{\rm ext})
=
V_3\,\mu_{\rm ext}\,\rho_i^2(\mu_{\rm ext}),
\label{eq:Qi_of_mu}
\end{equation}
where $\rho_i(\mu_{\rm ext})$ denotes the metastable homogeneous solution determined by the reduced grand-canonical potential.

There are then two different dynamical regimes. If charge exchange with the reservoir is inefficient on the tunnelling time scale, the chemical potential prepares the initial homogeneous state, but the semiclassical tunnelling event itself takes place at fixed total charge. In that case one first uses $\mu_{\rm ext}$ to determine the initial charge sector, 
$\mu_{\rm ext}
\;\longrightarrow\;
Q_i(\mu_{\rm ext}),
$
and then solves precisely the fixed-$Q$ problem developed in this paper with
$
Q=Q_i(\mu_{\rm ext}).$

The residual twist $\Delta\eta $ is still determined by the conservation of the charge, because it enforces that the bounce and the homogeneous reference configuration lie in the same charge sector. The suppression exponent for the decay of a homogeneous state prepared at chemical potential $\mu_{\rm ext}$ is therefore
\begin{equation}
B(\mu_{\rm ext})
=
S_{Q_i(\mu_{\rm ext}),\beta},
\label{eq:B_mu_from_fixed_Q}
\end{equation}
with
\begin{equation}
S_{Q,\beta}
=
S_E[\phi_b,\bar\phi_b]
-
S_E[\phi_i,\bar\phi_i]
+
\Delta\eta  Q .
\label{eq:FQbeta_recalled}
\end{equation}
There is no additional $-\beta\mu_{\rm ext}Q$ term  in
\eqref{eq:B_mu_from_fixed_Q}. Such term in the exponent comes from the factor $e^{\beta\mu_{\rm ext}Q}$, 
which appears only if one computes a grand-canonical average over different charge sectors. However, $S_{Q,\beta}$ is not the
grand-canonical statistical weight of the charge sector $Q$, but the
conditional tunnelling exponent once that sector has been fixed.
All extensive contributions, associated with the preparation of
that fixed-$Q$ state, have already been divided out in the normalization of
the decay probability giving $S_E[\phi_b,\bar\phi_b]
-
S_E[\phi_i,\bar\phi_i]$. In this regime $\mu_{\rm ext}$ is used only to
identify the charge of the initial homogeneous state,
$
Q=Q_i(\mu_{\rm ext}),
$
whereas the subsequent semiclassical transition is a fixed-charge process.
\footnote{This observation also explains how the fixed-$Q$ scans shown above can be reinterpreted as scans in the 
chemical potential of the initial state, provided the map between $\mu_{\rm ext}$ and $Q_i$ is 
one-to-one along the metastable branch. The precise relation is
$Q_i(\mu)/Q_{\max}
=
\mu\,\rho_i^2(\mu)/[
\mu_{\max}\,\rho_i^2(\mu_{\max})]$.
Only if $\rho_i(\mu)$ varies negligibly along the branch does this reduce to the simple identification $Q/Q_{\max}\simeq \mu/\mu_{\max}$.}

The opposite regime is one in which the system remains in efficient contact with a reservoir throughout the tunnelling event. Then the charge of the tunnelling configuration is not fixed. Instead, the chemical potential is fixed, and the charge of the saddle is an output. In such a genuinely fixed-$\mu$ calculation, one should not project onto a definite charge sector. Equivalently,
\begin{equation}
\text{fixed }Q:
\qquad
Q\ \text{fixed},\quad \Delta\eta \ \text{tuned},
\end{equation}
whereas
\begin{equation}
\text{fixed }\mu:
\qquad
\mu\ \text{fixed},\quad Q\ \text{output}.
\end{equation}
The same complex Euclidean framework remains the natural starting point, because the Wick rotation of a rotating charged state still turns the phase direction into a hyperbolic Euclidean direction. Thus, one should still allow $\phi$ and $\bar\phi$ to be independent Euclidean fields. What changes is the variational problem: the twist, or equivalently the Euclidean chemical potential, is prescribed by $\mu$, rather than being adjusted to enforce a target charge.

The corresponding exponent would be a grand-canonical action difference,
\begin{equation}
B_{\mu,\beta}
=
S_{E,\mu}[\phi_b,\bar\phi_b]
-
S_{E,\mu}[\phi_{i},\bar\phi_{i}] ,
\label{eq:B_mu_grand_canonical}
\end{equation}
where $(\phi_{i},\bar\phi_{i})$ is the homogeneous
metastable configuration at the prescribed value of $\mu$, while
$(\phi_b,\bar\phi_b)$ is the nontrivial saddle of the same
grand-canonical Euclidean functional. Operationally, one would fix
$\mu$ from the start and solve the Euclidean PDE problem with
the corresponding chemical-potential twist, or equivalently, with the
shifted Euclidean derivative $
\partial_\tau \longrightarrow \partial_\tau-\mu
$
in the action. The bounce is then found by extremising
$S_{E,\mu}[\phi,\bar\phi]$ directly, subject to the appropriate
turning-slice and asymptotic false-state boundary conditions. In contrast
with the fixed-$Q$ computation, there is no need to enforce the conservation of the charge. The charge carried by the saddle is instead an output of the solution,
\begin{equation}
Q_b(\mu)
=
Q[\phi_b,\bar\phi_b],
\end{equation}
and need not be imposed to coincide with a prescribed value beforehand. This is why \eqref{eq:B_mu_grand_canonical} is not the same object as the
fixed-charge Legendre-transformed exponent in \eqref{eq:FQbeta_recalled}. The term $\Delta\eta  Q$ appears only in the projected fixed-charge problem,
where the twist is varied so as to impose a given charge sector. In a
genuinely fixed-$\mu$ calculation the chemical potential is prescribed,
the twist is fixed, and the decay exponent is computed directly from the
difference of grand-canonical Euclidean actions. This is a closely related,
but distinct, open-system tunnelling problem.

Finally, let us relate this discussion to the usual grand-canonical decomposition. Formally,
\begin{equation}
Z(\beta,\mu)
=
\mathrm{Tr}\,e^{-\beta(H-\mu Q)}
=
\sum_Q e^{\beta\mu Q} Z_Q(\beta).
\label{eq:grand_canonical_decomposition}
\end{equation}
This expression is useful when one wants an ensemble average over charge sectors. However, for a specified homogeneous initial state prepared at a given $\mu$, the thermodynamic distribution is sharply peaked around $Q_i(\mu)$ in the large-volume limit. Therefore, if charge exchange is slow during nucleation, the relevant exponent is \eqref{eq:B_mu_from_fixed_Q}, not a minimization over arbitrary charge sectors. A sum or saddle over $Q$ is appropriate only if one is computing the decay rate of an unconditioned grand-canonical ensemble rather than the decay of a specified homogeneous charged background.

\section{Conclusions and outlook}
\label{sec:conclusion}

We have developed a systematic formulation of semiclassical bubble nucleation at fixed conserved global charge $Q$ out of a homogeneous charged medium, and we have solved the associated Euclidean saddle problem explicitly. The main lesson is that finite charge does not merely deform the standard Coleman analysis quantitatively: it changes the structure of the tunnelling problem itself. Once the decay is formulated in a definite charge sector, the path integral must be charge projected, the Euclidean-time boundary conditions become twisted, and the dominant saddle is generically complex in the original field variables. A real and computation-ready formulation is nevertheless obtained by treating, in the $U(1)$ case, $\phi$ and $\bar\phi$ as independent Euclidean fields and matching them on a turning slice from which the Minkowski evolution is reconstructed.

The conceptual ingredients of the problem were already present in earlier literature, especially in Lee's analysis of tunnelling with global charge and in fixed-charge wormhole analyses \cite{Lee:1994bza,Lee:1988ge,Coleman:1989zu,Giddings:1988cx,Giddings:1988wv}. However, the full problem had not been carried through to the end in the context relevant here. In particular, what was missing was a systematic and numerically explicit treatment of bubble nucleation from a homogeneous charged medium: a formulation in which the fixed-$Q$ exponent is defined unambiguously relative to the homogeneous charged reference state, the charged Euclidean saddle is solved as a PDE problem, the neutral limit is recovered smoothly, and the post-nucleation real-time evolution is tracked in a way consistent with the fixed-charge construction. This is what the present work provides, and, for reproducibility, all numerical material and calculations we used are available on the GitHub page \href{https://github.com/GiulioBarni/Qubble}{\faGithub\ Qubble}.

Our analysis clarifies the physical interpretation of the charged saddle. At small charge, the dominant effect is a controlled deformation of the neutral bounce together with the reduction of the critical radius anticipated by the thin-wall estimate. At larger charge, the Euclidean saddle departs appreciably from exact $O(4)$ symmetry. Having a nonzero charge lowers the tunnelling action, facilitating vacuum decay.
The posterior real-time evolution is not a trivial continuation of the neutral problem with shifted parameters. Rather, the phase sector remains dynamical, carries the conserved charge, rearranges it across the expanding interface, and contributes significantly to the wall energy.

One of the main physical outcomes of this work is this real-time effect. The late-time Minkowski evolution is an essential prediction of fixed-charge tunnelling. In the examples studied here, the charge-flow sector makes the wall energy grow linearly with the bubble radius, and this in turn leads to a terminal wall velocity smaller than one even at zero temperature. This is a genuinely new feature absent in neutral vacuum decay, and it illustrates that finite charge modifies both the tunnelling rate and the subsequent bubble dynamics. In particular, since the bubble-wall velocity is a key ingredient in determining the efficiency and spectral shape of the resulting gravitational-wave signal, such a subluminal terminal regime can significantly alter the predicted gravitational-wave phenomenology. 

An interesting direction opened by the charged dynamics studied here concerns gravitational-wave production already at zero temperature. In the standard neutral vacuum case, the collision of two bubbles is strongly constrained by the symmetry of the two-bubble configuration: for identical bubbles nucleated in vacuum, the dynamics retains an $SO(1,2)$ symmetry, which severely restricts the possible transverse-traceless components of the stress tensor. In the charged case, however, the finite-density background selects a preferred frame and the Lorentzian evolution involves a non-trivial redistribution of phase gradients and charge density. We therefore expect the charged two-bubble configuration to break the symmetry structure of the neutral vacuum problem in a way that can provide an intrinsically anisotropic and time-dependent source. This suggests that bubble collisions in a charged medium may generate a gravitational-wave signal even in the absence of a thermal plasma. A quantitative computation of the transverse-traceless stress tensor and of the resulting spectrum is left for future work.

More broadly, the present paper should be viewed as a first concrete step
toward quantitative nucleation estimates in systems at finite conserved
density. The fixed-$Q$ bounce is the relevant microscopic object whenever
the tunnelling event is faster than the processes that exchange charge with
the environment. This remains true even if the macroscopic state is most
naturally prepared or labelled by a chemical potential. In that case the
chemical potential selects the initial homogeneous charged state, and hence
the corresponding charge $Q_i(\mu)$, while the subsequent semiclassical
transition proceeds in that fixed charge sector. The decay exponent of such
a finite-$\mu$-prepared state is therefore obtained from the same
fixed-$Q$ saddle, evaluated at $Q=Q_i(\mu)$.

Our discussion also clarifies the distinction between different uses of a
chemical potential in tunnelling problems. If the chemical potential belongs
to spectator or integrated-out degrees of freedom, its effect is encoded in
the finite-density effective potential for the tunnelling field. If instead
it is conjugate to the charge carried by the tunnelling field itself, the
same complex Euclidean structure encountered here is unavoidable: the
charged homogeneous state rotates in the internal $U(1)$ direction, and
after Wick rotation this becomes a hyperbolic Euclidean phase. A genuinely
fixed-$\mu$ open-system calculation, in which charge can be exchanged with
a reservoir during the tunnelling event, would require a modified
variational problem with $\mu$ fixed and the charge of the saddle left as
an output. Although not identical to the fixed-$Q$ problem solved here, this formulation is a close variant of it and should be straightforward to implement within the same numerical framework, opening the way to quantitative fixed-$\mu$ computations.

This distinction provides a useful way of organizing possible finite-density
applications. Dense QCD, for instance, features both possibilities. In
effective descriptions of the chiral/deconfinement transition, the relevant
collective variables are usually neutral under baryon number: a meson-like
chiral field, schematically $\sigma\sim\bar q q$, and, in PNJL-type
descriptions, a Polyakov-loop sector \cite{Fukushima:2003fw,Ratti:2005jh,
Schaefer:2004en}. The baryon or quark chemical potential is then carried by
the quark degrees of freedom and affects nucleation through the
finite-density effective potential obtained after integrating them out. This
realizes the spectator-density case discussed above: the transition is
certainly a finite-density first-order transition, but the tunnelling
collective coordinate is not itself charged under $U(1)_B$. A more direct
target for the fixed-charge formalism is instead a transition from hadronic
matter to a color-superconducting phase. In an effective description of such
a phase, the relevant order parameter is diquark-like,
$\Delta\sim\langle q q\rangle$, and carries baryon number,
$B=2/3$ for $B_q=1/3$. Its phase is therefore associated with the
breaking of $U(1)_B$, most notably in the CFL phase, which is a superfluid
phase of dense quark matter \cite{Alford:1998mk,Alford:2007xm,Buballa:2003qv}.
In this setting the conserved density is carried by the tunnelling order
parameter itself, so the twisted Euclidean boundary conditions, complex
fixed-charge saddle, and charge-conserving real-time evolution developed in
this work become directly relevant. A realistic QCD implementation would
still require extending the present single-field construction to multifield
dynamics, colour gauge constraints, and electric/colour neutrality conditions.

To conclude, there are several natural directions for further work. On the formal side,
the next priority would be the computation of fluctuation determinants and
prefactors around the fixed-$Q$ saddle, together with a systematic
treatment of negative and zero modes. It would also be important to develop
the genuinely fixed-$\mu$ formulation mentioned above, replacing the
fixed-charge by fixed-chemical-potential boundary conditions or
an equivalent Euclidean action with a chemical-potential covariant
derivative. On the phenomenological side, natural extensions include
multifield theories, locally conserved or gauge charges, spectator charged
sectors integrated into finite-density effective potentials, and settings
in which hydrodynamic transport and finite-density backgrounds play a
central role.

A further direction we are exploring is a reformulation of fixed-charge tunnelling in the Minkowski-space most probable escape path language of Bitar and Chang \cite{Bitar:1978vx,Bitar:1977wy}. Such a formulation could provide a complementary real-time understanding of the charged tunnelling trajectory, make the role of the constrained phase direction more transparent, and help clarify the relation between the Euclidean fixed-$Q$ saddle and the corresponding escape path in complexified field space.

These developments are necessary if one wants to turn the fixed-charge
bounce from a proof of principle into a quantitative tool for phase
conversion in realistic dense systems. In that sense, the present work
should be read as a first fully worked-out benchmark. It establishes that
the fixed-$Q$ tunnelling problem can be posed sharply, solved numerically,
and interpreted physically all the way from the Euclidean saddle to the
late-time Minkowski bubble. It also shows how this fixed-charge object
interfaces with finite-temperature and finite-chemical-potential
descriptions: finite $T$ changes the Euclidean time extent and the
effective potential, while finite $\mu$ either dresses the tunnelling
potential through spectator-charged sectors or selects the charge of the
initial rotating state. This, in our view, is the necessary starting point
for any future quantitative treatment of vacuum decay and first-order phase
conversion at finite conserved density.

\paragraph{Acknowledgements.} We thank A. Azatov, J. Serra, and T. Steingasser for discussions during different stages of this work. GB acknowledges support from the Spanish Research Agency (Agencia Estatal de Investigación, MCIN/AEI/10.13039/501100011033) via the IFT Severo Ochoa centre of Excellence grant CEX2020-001007-S. In addition, GB is supported by the grant CNS2023-145069 funded by MICIU/AEI/10.13039/501100011033 and by the European Union NextGenerationEU/PRTR. J.R.E. acknowledges CERN TH Department for hospitality while this research was being carried out. The work of J.R.E. has been supported by the IFT Centro de Excelencia Severo Ochoa CEX2020-001007-S and by PID2022-142545NB-C22 funded by MCIN/AEI/10.13039/ 501100011033 and by “ERDF A way of making Europe”.

\appendix

\section{Late-time estimate for the phase-sector contribution}
\label{app:late_time_estimate}

In this appendix we clarify the role of the two phase-sector contributions 
\begin{equation}
e_{\theta,t}=\frac12 \rho^2 \dot\theta^2,
\qquad
e_{\theta,r}=\frac12 \rho^2(\partial_r\theta)^2,
\end{equation}
in the late-time Minkowski evolution.
The main point is that these two terms have different physical meanings. The
temporal term behaves as a bulk rotational energy and helps the conversion of the initial state to
the inner phase, whereas the radial-gradient term is localised near the wall and
provides a growing positive cost that opposes the wall acceleration.

\begin{figure}[t]
    \centering
    \includegraphics[width=\textwidth]{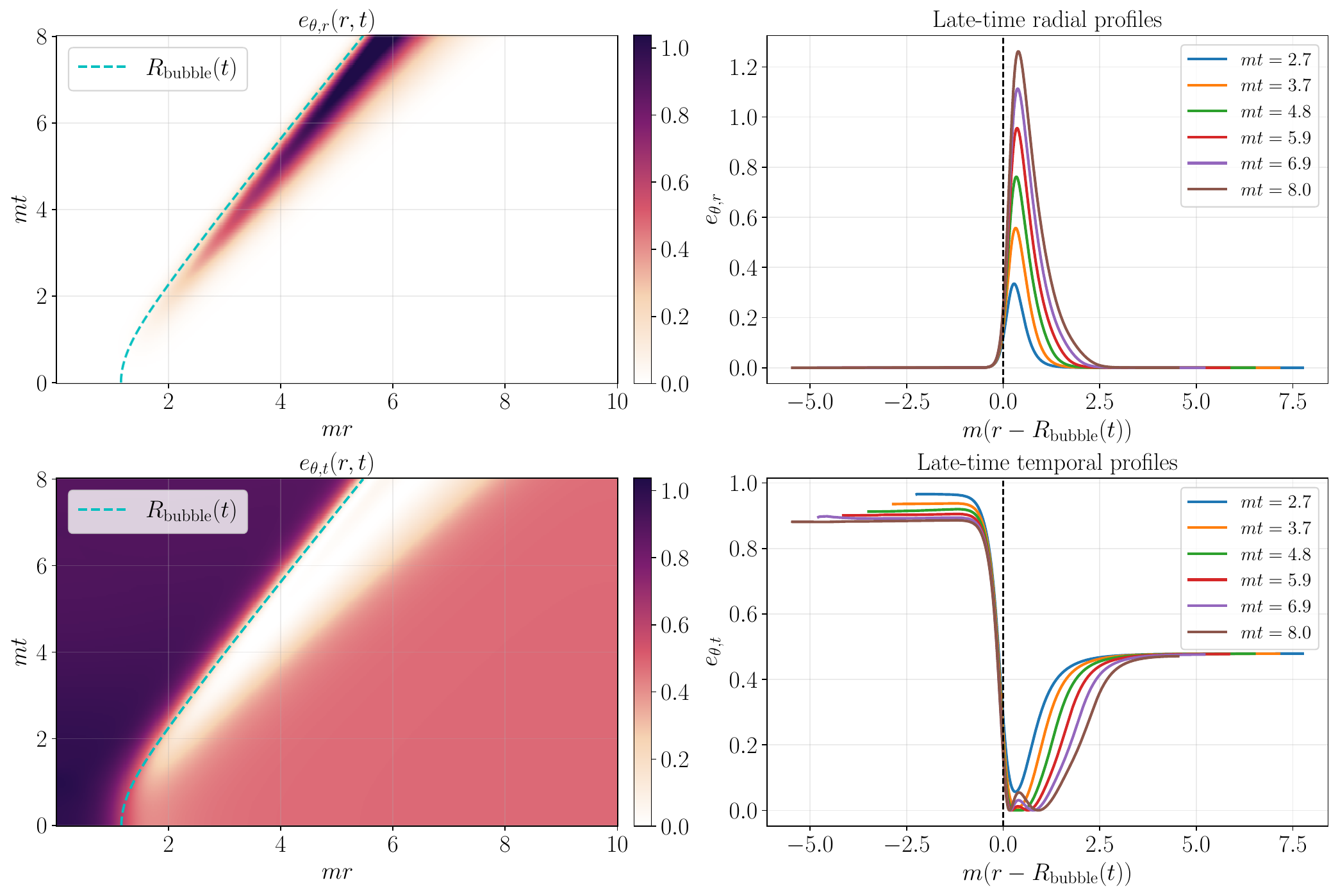}
    \caption{
    Phase-sector energy densities during the Minkowski evolution. 
Top row: radial phase-gradient energy 
$e_{\theta,r}$, shown as a spacetime map and as late-time profiles versus $r-R_{\rm bubble}(t)$. It remains localised in a thin shell near the wall, while its peak grows with time. 
Bottom row: temporal phase energy 
$e_{\theta,t}$. This term has a broad plateau structure, reflecting the different rotational energies of the two homogeneous phases, and therefore contributes to the effective bulk-driven expansion rather than to a localised wall cost opposing it.
    }
    \label{fig:minkowski_phase_energy}
\end{figure}

This separation between the two phase-sector contributions is visible directly in
Fig.~\ref{fig:minkowski_phase_energy}: $e_{\theta,t}$ forms broad plateaux
associated with the different homogeneous rotational energies of the two phases,
whereas $e_{\theta,r}$ remains localised in a narrow shell slightly in front of
the moving wall, with a peak that grows during the evolution.

At late times the bubble profile is well approximated by two homogeneous
plateaus separated by a thin wall. We denote the inner and outer plateau values
by
\begin{equation}
(\rho_-,\omega_-),
\qquad
(\rho_+,\omega_+),
\qquad
\omega_\pm\equiv \dot\theta_\pm,
\qquad
\omega_-<\omega_+,
\end{equation}
and the bubble radius and Lorentz factor by $R(t)$ and $\gamma(t)$.

\paragraph{Temporal phase term: a bulk gain.}

In a locally homogeneous region the charge density is
\begin{equation}
q=\rho^2\omega.
\end{equation}
Therefore
\begin{equation}
e_{\theta,t}
=
\frac12 \rho^2\omega^2
=
\frac12 q\,\omega
=
\frac{q^2}{2\rho^2}.
\label{eq:app_temporal_phase_density}
\end{equation}
This makes the physical interpretation transparent: at fixed charge density,
the rotational energy is smaller in the phase with larger $\rho$. Since in
the solutions considered here $\rho_->\rho_+$, charge is stored more cheaply
inside the bubble.

Assuming that the two plateaus carry approximately the same local charge
density $q$, one finds
\begin{equation}
e_{\theta,t}^{(+)}-e_{\theta,t}^{(-)}
=
\frac12 q(\omega_+-\omega_-)
=
\frac12 q\,\Delta\omega,
\qquad
\Delta\omega\equiv\omega_+-\omega_->0.
\label{eq:app_temporal_bulk_gain_density}
\end{equation}
Thus, replacing a volume of the outer phase with the inner phase gives a positive
bulk gain
\begin{equation}
E_{{\rm gain},\theta,t}(R)
=
\frac{4\pi}{3}R^3
\left(e_{\theta,t}^{(+)}-e_{\theta,t}^{(-)}\right)
=
\frac{2\pi}{3}q\,\Delta\omega\,R^3.
\label{eq:app_temporal_bulk_gain}
\end{equation}
Equivalently, if $Q_R=(4\pi/3)R^3 q$ is the charge contained in the converted
volume,
\begin{equation}
E_{{\rm gain},\theta,t}(R)
=
\frac12\,\Delta\omega\,Q_R.
\end{equation}
The temporal phase term is therefore linear in the charge contained in the
converted region. It contributes to the effective volume-driven expansion, not to the
wall cost opposing it.

\paragraph{Radial phase term: a localised growing wall cost.}

The radial phase-gradient contribution is
\begin{equation}
e_{\theta,r}
=
\frac12 \rho^2(\partial_r\theta)^2.
\end{equation}
The numerical evolution shows that this term remains localised in a narrow shell
attached to the wall, slightly in front of the moving interface. Its support does
not grow macroscopically with $R$. We therefore write
\begin{equation}
E_{\theta,r}^{\rm shell}(R)
=
4\pi
\int_{R-\ell}^{R+\ell}
dr\,r^2 e_{\theta,r}(r,t),
\label{eq:app_Etheta_r_shell_def}
\end{equation}
where $\ell$ is a small shell width. Since $\ell\ll R$, this becomes
\begin{equation}
E_{\theta,r}^{\rm shell}(R)
\simeq
4\pi R^2
\int_{-\ell}^{+\ell}
ds\, e_{\theta,r}(R+s,t)
\equiv
4\pi R^2 \ell\,\bar e_{\theta,r}(R).
\label{eq:app_Etheta_r_shell}
\end{equation}
If $\bar e_{\theta,r}$ were constant, this would simply renormalize the usual
surface tension. Instead, the simulations show that the support remains narrow
while the peak height grows approximately linearly with time, and therefore
approximately linearly with $R$ once the wall velocity has nearly saturated:
\begin{equation}
\bar e_{\theta,r}(R)
\simeq
e_0+cR.
\label{eq:app_ebar_linear}
\end{equation}
Consequently,
\begin{equation}
E_{\theta,r}^{\rm shell}(R)
\simeq
4\pi\ell e_0 R^2
+
4\pi\ell c R^3.
\label{eq:app_Etheta_r_cubic}
\end{equation}
Thus, a shell-localised contribution  becomes effectively
volume-like after integration, not because its width grows as $R$, but because its amplitude grows with $R$.

The corresponding rest-frame contribution to the effective surface energy is
\begin{equation}
\sigma_{\theta,r}^{\rm rest}(R)
=
\frac{E_{\theta,r}^{\rm shell}(R)}{4\pi R^2\gamma(R)}
\simeq
\frac{\ell e_0}{\gamma_\infty}
+
\frac{\ell c}{\gamma_\infty}R,
\label{eq:app_sigma_theta_r_linear}
\end{equation}
where in the last step we used $\gamma(R)\to\gamma_\infty$. This gives the
observed approximately linear growth of the effective wall energy.

\paragraph{Net force and terminal velocity.}

The large-$R$ energy balance can be written schematically as
\begin{equation}
\Delta E(R)
=
E_{\rm wall}(R)
-
\frac{4\pi}{3}\Delta P_{\rm eff}\,R^3,
\label{eq:app_energy_balance}
\end{equation}
where $E_{\rm wall}$ includes the ordinary wall tension, the modulus-gradient
contribution, the phase-gradient contribution, and the remaining
bulk-subtracted wall terms. The effective driving pressure is
\begin{equation}
\Delta P_{\rm eff}
=
\left[V_+ + e_{\theta,t}^{(+)}\right]
-
\left[V_- + e_{\theta,t}^{(-)}\right].
\label{eq:app_deltaP_eff}
\end{equation}
Equivalently, the temporal phase contribution appears in
$\Delta P_{\rm eff}$ through the positive gain density in
Eq.~\eqref{eq:app_temporal_bulk_gain_density}.

The net outward force per unit area is
\begin{equation}
F_{\rm net}(R)
=
-\frac{1}{4\pi R^2}\frac{d\Delta E}{dR}
=
\Delta P_{\rm eff}
-
\frac{1}{4\pi R^2}\frac{dE_{\rm wall}}{dR}.
\label{eq:app_Fnet}
\end{equation}
A terminal wall velocity requires
\begin{equation}
F_{\rm net}(R)\longrightarrow 0
\qquad
(R\to\infty).
\end{equation}

To conclude, the late-time picture is thus the following. The temporal phase energy lowers
the bulk rotational cost of the inner phase and therefore helps the bubble
expand. The radial phase-gradient energy, instead, is localised near the wall, but
its amplitude grows with the radius. Its integrated contribution becomes
approximately cubic in $R$, producing an effective pressure cost that can
balance $\Delta P_{\rm eff}$ and drive the system toward a subluminal terminal
velocity.

\bibliographystyle{JHEP}
{
\bibliography{biblio}}

\end{document}